\documentclass{aa}  
\usepackage{graphicx}
\usepackage{txfonts}
\usepackage[colorlinks=true,allcolors=blue]{hyperref}
\usepackage{color}
\usepackage{multirow}
\newcommand{\Lext}{$\mathcal{L}_{\rm EXT}$}

\usepackage{xcolor}

\begin{document} 
\title{The SRG/eROSITA All-Sky Survey}
\subtitle{Tracing the Large-Scale Structure with a clustering study of galaxy clusters
}
\author{
R. Seppi\inst{1, 2}\thanks{E-mail: riccardo.seppi@unige.ch} \and
J. Comparat\inst{1} \and
V. Ghirardini\inst{1} \and
C. Garrel\inst{1} \and
E. Artis\inst{1} \and
A.~G.~S\'anchez\inst{1} \and
A. Liu\inst{1} \and
N. Clerc\inst{3} \and
E. Bulbul\inst{1} \and
S. Grandis\inst{4} \and
M. Kluge\inst{1} \and
T. H. Reiprich\inst{5} \and
A. Merloni\inst{1} \and
X. Zhang\inst{1} \and
Y. E. Bahar\inst{1} \and
S. Shreeram\inst{1} \and
J. Sanders\inst{1} \and
M. Ramos-Ceja\inst{1} \and
M. Krumpe\inst{6}
}
\institute{
Max-Planck-Institut f\"{u}r extraterrestrische Physik (MPE), Giessenbachstraße 1, D-85748 Garching bei M\"unchen, Germany \and
Department of Astronomy, University of Geneva, Ch. d’Ecogia 16, CH-1290 Versoix, Switzerland \and
IRAP, Université de Toulouse, CNRS, UPS, CNES, Toulouse, France \and
Universit\"at Innsbruck, Institut für Astro- und Teilchenphysik, Technikerstr. 25/8, 6020 Innsbruck, Austria \and
Argelander-Institut f\"{u}r Astronomie (AIfA), Universit\"{a}t Bonn, Auf dem H\"{u}gel 71, 53121, Bonn, Germany \and
Leibniz-Institut für Astrophysik Potsdam, An der Sternwarte 16, 14482 Potsdam, Germany
}

\date{Accepted XXX. Received YYY; in original form ZZZ}

\titlerunning{The clustering of eRASS1 galaxy clusters}
\authorrunning{Seppi et al.}

\abstract{
The spatial distribution of galaxy clusters provides a reliable tracer of the large-scale distribution of matter in the Universe. The clustering signal depends on intrinsic cluster properties and cosmological parameters.}
{
%
The ability of eROSITA onboard Spectrum-Roentgen-Gamma (SRG) to discover galaxy clusters allows probing the association of extended X-ray emission to dark matter haloes. We aim to measure the projected two-point correlation function to study the occupation of dark matter halos by clusters and groups detected by the first eROSITA all-sky survey (eRASS1).}
{
%
We create five volume-limited samples probing clusters with different redshift and X-ray luminosity. We interpret the correlation function with halo occupation distribution (HOD) and halo abundance matching (HAM) models. 
We simultaneously fit cosmological parameters and halo bias of a flux-limited sample of 6493 clusters with purity $> 96\%$.}
{%
We obtain a detailed view of the halo occupation for eRASS1 clusters. The fainter population at low redshift (S0: $\overline{L_X}=4.63\times 10^{43}\, {\rm erg/s}$, $0.1<z<0.2$) is the least biased compared to dark matter, with $b=2.95\pm0.21$. The brightest clusters up to higher redshift (S4: $\overline{L_X}=1.77\times 10^{44}\, {\rm erg/s}$ , $0.1<z<0.6$) exhibit a higher bias $b=4.34\pm0.62$. Satellite groups are rare, with a satellite fraction $<14.9\%\, (8.1)$ for the S0 (S4) sample. We combine the HOD prediction with a HAM procedure to constrain the scaling relation between $L_{\rm X}$ and mass in a new way and find a scatter of $\langle \sigma_{\rm Lx} \rangle = 0.36$. We obtain cosmological constraints for the physical cold dark matter density $\omega_{\rm c}=0.12^{+0.03}_{-0.02}$ and an average halo bias $b=3.63^{+1.02}_{-0.85}$. 
}
{
We model the clustering of galaxy clusters with a HOD approach for the first time, paving the way for future studies combining eROSITA with 4MOST, SDSS, \textit{Euclid}, \textit{Rubin}, and DESI to unravel the cluster distribution in the Universe.
}

\maketitle

\section{Introduction}

The evolution of the large-scale structure (LSS) of the Universe encodes crucial information about the nature of dark matter and dark energy. Cosmic structures are the result of a hierarchical process of structure formation, starting from tiny density perturbations in the early Universe that collapsed under the action of gravity into dark matter haloes hosting galaxies and clusters we observe nowadays \citep[][]{Lacey_Cole1993MNRAS.262..627L, Coil2013_LSS}. \\
Clusters of galaxies are hosted by the most massive dark matter haloes located at the nodes of the LSS \citep[][]{Kaiser1987MNRAS.227....1K}. Therefore, they provide an excellent tracer of the overall distribution of matter in the Universe and can probe the growth of cosmic structures \citep[][]{Allen2011, Kravtsov2012ARA&ABorgani, Pratt2019SSRv..215...25P, Clerc2022arXiv220311906C_review}. Their abundance as a function of mass and redshift (i.e., the measure of the halo mass function) provides powerful constraints on the total amount of matter in the Universe and the amplitude of density fluctuations \citep[][]{Mantz2015cosmology, Bocquet2019spt_cosmo, IderChitham2020MNRAS.499.4768I, Finoguenov2020A&A...638A.114F, Costanzi2021desclucosmo, Garrel2022A&A...663A...3G, Lesci2022A&A...659A..88L}.

The spatial distribution of clusters, quantified in summary statistics such as the two-point correlation function, encodes additional cosmological information  \citep[][]{Marulli2018A&A...620A...1M}. The correlation function expresses the excess probability of finding a couple of objects at a given separation compared to a randomly distributed sample. \\
Several phenomena impact the measure and modelling of clustering. Redshift space distortions \citep[RSD,][]{Scoccimarro1999ApJ...517..531S, Percival2009MNRAS_RSD} introduce an anisotropic warping of the correlation function when the comoving distance is estimated from redshift. Such an effect is due to peculiar velocities and intrinsic motions, introducing an additional component to the redshift of an extra-galactic source on top of the Hubble flow. For galaxy clusters, most of the impact on RSD comes from the Kaiser effect \citep[][]{Kaiser1987MNRAS.227....1K}, which squeezes the correlation function along the line of sight due to large-scale motions towards large overdensity regions. On smaller scales, random motions of virialized objects stretch the density fields, reducing the clustering power along the line of sight and causing the Finger of God effect \citep[][]{Jackson1972MNRAS_fingerofgod}. One can minimize the impact of RSD on the correlation function by marginalizing clustering along the line of sight \citep[][]{DavisPeebles1983ApJ_wprp, Zehavi2011ApJ_SDSSclustering, More2023}. \\
On the one hand, most clustering studies in the literature involve individual galaxies, thanks to large catalogues provided by recent galaxy surveys, such as the Dark Energy Survey \citep[DES,][]{Abbott2022PhRvD_DESY3}, the Sloan Digital Sky Survey \citep[SDSS,][]{Alam2021PhRvD.103h3533A_SDSS}, and the  Kilo-Degree Survey \citep[KiDS,][]{Heymans2021A&A_KIDScosmo}. These large surveys allow a detailed understanding of the connection between the large-scale structure and its tracers. The halo occupation distribution \citep[HOD, see][for an example]{Berlind2002ApJHOD} framework tackles this questions directly. It describes how central and satellite objects populate dark matter haloes of different masses \citep[][]{Peacock2000MNRAS.318.1144P}. Given a set of haloes generated by a cosmological model, the HOD encodes the distribution of a given tracer (e.g., galaxies) within such haloes. Therefore, it contains precious information about galaxy formation and baryonic gas properties \citep[][]{Zheng2005ApJ_HOD}. In the literature, it provided a detailed view of the galaxy distribution in the LSS \citep[][]{Zehavi2005ApJ...630....1Z, Zheng2007ApJ_HOD, Zehavi2011ApJ_SDSSclustering, Zhou2021MNRAS_clusteringDESI_LRG, Rocher2023_HODdesi}. The HOD framework has also been applied to active galactic nuclei (AGN), combining X-ray and optical data \citep[][]{Krumpe2010ApJ_agn_clustering_rosat, Krumpe2015ApJHOD, Krumpe2023ApJ...952..109K, Comparat2023A&Aagnclustering}. Recent developments with emulators allow flexible HOD modeling \citep[][]{Nishimichi2019ApJ...884...29N, Salcedo2022MNRAS_HODemu}.\\ 
On the other hand, clustering studies with galaxy clusters present several advantages. Non-linear physics at small scales is a less dominant effect compared to galaxy studies, which reduces modelling uncertainties due to the non-linear matter power spectrum, as well as the impact of baryons on scales larger than about 10 Mpc \citep[][]{Hamilton1992ApJ_multipoles, Marulli2017A&A...599A.106M}. In addition, the ability to probe cluster masses, for example via weak lensing \citep[see e.g.,][]{Bulbul2019ApJ...871...50B_scalingrel, umetsu2020clustergalaxy, Schrabback2021MNRAS_WL, Grandis2021MNRAS_WLmass, Chiu2022A&A...661A..11C}, allows us to directly model the bias for a given cluster sample compared to the clustering of dark matter, according to the peak background split \citep[][]{Bardeen1986ApJBBKS, MoWhite1996MNRAS_clustering, Sheth1999MNRAS_LSbias}. More recently, numerical simulations allowed the development of precise large-scale halo bias models \citep[][]{Tinker2010ApJ_bias, Bhattacharya2011, Comparat2017}.
As a result, cluster clustering studies provide competitive cosmological constraints on their own \citep[][]{Borgani1999MNRAS_abell_clustering, Moscardini2000MNRAS_RASS, Balaguera2011MNRAS_reflexPk, Hong2016ApJ_BAO_clu_sdss3, Veropalumbo2016MNRAS.458.1909V_BAO, Marulli2018A&A...620A...1M, Marulli2021ApJ...920...13M, Lindholm2021A&A...646A...8L, Ingoglia2022MNRAS_KidsDR3_halobias, Lesci2022A&A_clustering, Lesci2023A&A_planckclustering, Romanello2023arXiv_KiDS_clustering} and improve constraining power from cluster counts alone \citep[][]{Mana2013MNRAS.434..684M, Sartoris2016MNRAS_cosmopred, Pillepich2018, Garrel2022A&A...663A...3G}. 

Thanks to the eROSITA (extended ROentgen Survey with an Imaging Telescope Array) X-ray telescope onboard Spectrum-Roentgen-Gamma \citep[SRG,][]{Merloni2012, Predehl2021A&Aerosita}, a large number of clusters and groups are being discovered. 
eROSITA conducted its first all sky survey between December 2019 and June 2020 (eRASS1). 
The sky is split between the western and eastern galactic hemispheres, respectively owned by the German (eROSITA\_DE) and Russian (eROSITA\_RU) consortia. In this work, we focus on the eROSITA\_DE sky. The full eRASS1 source catalogue and the cluster catalogue are presented in \citet{Merloni2024A&A_erass1cat}, \citet{Bulbul2024}. 12,247 cluster candidates in eRASS1 have been confirmed by an optical follow-up as described in \citet{Kluge2024}. The main cosmological results using cluster counts as a probe are presented in \citet{Ghirardini2024}. Other cosmological results are presented by \citet{Garrel2024}, with a photon-based exploration of the angular power spectrum, and \citet{Artis2024}, with a study of alternative theories of gravity. A collection of superclusters is presented by \citet{Liu2024supercluster}.

In this work, we study the clustering properties of the eRASS1 cluster sample in the real space using the projected correlation function. We interpret our measurements for different eRASS1 subsamples with a halo occupation distribution approach. To our knowledge, this is the first attempt to model the X-ray selected cluster distribution in the LSS with HOD. We investigate the possibility of massive dark matter halos to host satellite substructures. In fact, eROSITA may start probing a population of low mass groups that are satellites to the largest dark matter haloes (>10$^{15}$ M$_\odot$). This is also expected from models of the subhalo mass function \citep[][]{Dolag2009MNRAS_subhalomf, Klypin2011ApJ_Bolshoi}. \\
We separately assess the cosmological implications of our measurements, focusing mostly on the cosmological parameters affecting the amplitude of the matter power spectrum \citep[][]{Sanchez2022MNRAS_evomapping}.

We organize this article as follows. In Sect. \ref{sec:data}, we present the eRASS1 cluster sample, its selection, and the construction of the random catalogues. In Sect. \ref{sec:method}, we describe how we measured the projected two-point correlation function and the models to constrain HOD and cosmological parameters. In Sect. \ref{sec:results}, we present our main findings, including a comparison of the best-fit HOD to a prediction from a halo abundance matching (HAM) procedure. Finally, in Sect. \ref{sec:conclusions}, we discuss our results and future prospects. \\
In this paper we use $\log$ as the base 10 logarithm. The X-ray luminosity is measured within $R_{\rm 500c}$\footnote{the radius enclosing a total mass equal to 500 times the critical density of the Universe at the cluster's redshift.} in the 0.2--2.3 keV band \citep[see][for details]{Bulbul2024}{}{}. We assume a \citet[][]{Planck2020A&A...641A...6P} cosmology, unless otherwise stated, with $\Omega_{\rm M}$=0.3112, $\sigma_{\rm 8}$=0.8102, and $h$=0.6774. 

\section{Data}
\label{sec:data}

\begin{figure}
    \centering
    \includegraphics[width=\columnwidth]{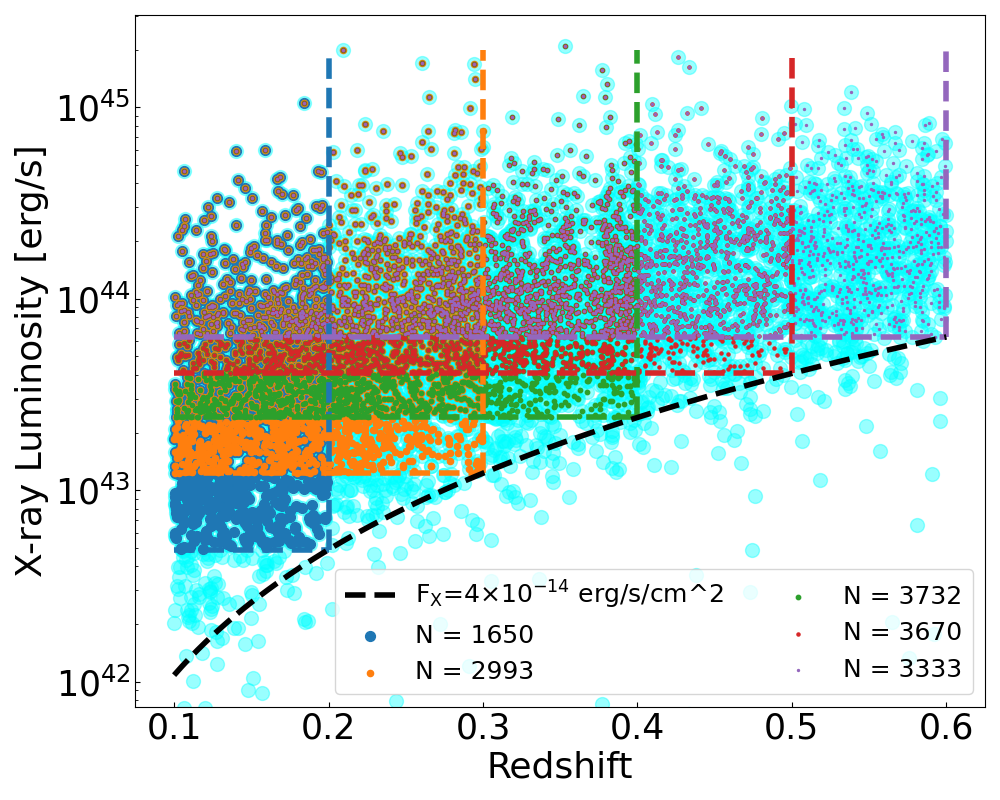}
    \caption{Volume-limited selection of the eRASS1 cluster sample for our clustering study as a function of X-ray luminosity and redshift. The full sample is shown in cyan, the five subsamples are denoted in blue, orange, green, red, and pink.}
    \label{fig:Lx_z_volume_limited}
\end{figure}

We use the clusters of galaxies detected during the first eROSITA all-sky survey in the Western galactic half of the sky \citep{Bulbul2024, Kluge2024}. 
We describe the treatment of the eROSITA data and the creation of the random catalogues in this section. The full eRASS1 source catalogue is presented in \citet{Merloni2024A&A_erass1cat}. We refer the reader to \citet[][]{Brunner2022_efedscat} for more details on the data processing and the detection pipeline with the eROSITA Science Analysis Software System (eSASS).

\subsection{eROSITA}
\label{subsec:erosita_data}
We start from the primary eRASS1 galaxy clusters and groups sample. Such catalogue has been cleaned from split sources as described in \citet{Bulbul2024}. We verified that the fraction of removed sources is similar to the one obtained from the eRASS1 digital twin \citep[][]{Seppi2022A&A_eRASS1sim} by accounting for multiple detections associated with the same cluster as well as secondary sources contaminating the detected clusters in the simulation.
We require a homogeneous optical coverage and further select clusters within the Legacy Survey DR10-South\footnote{\texttt{\url{https://www.legacysurvey.org/dr10}}} footprint. This excludes a small area of 462 deg$^2$ in the northern sky. Our final samples cover 12\,791 deg$^2$. A key property of the sample is the extent likelihood (\Lext), which encodes the probability for a given source to be extended (i.e., a cluster) compared to a point source. To maximize completeness, a low cut of \Lext $>3$ was chosen for eRASS1 \citep{Bulbul2024}. Accurate simulations showed that cluster samples are strongly contaminated by AGN at low values of extent likelihood \citep[][]{Seppi2022A&A_eRASS1sim}. \citet{Kluge2024} estimated the probability for each source to be a contaminant ($P_{\rm cont}$) via a mixture model. We obtain a purer cluster sample by further discarding sources with $P_{\rm cont}>0.5$. This strongly reduces the contamination due to bright AGN, making our sample >95$\%$ pure \citep{Bulbul2024}. We additionally select clusters with photometric redshift between 0.1 and 0.6, where the photo-z has $<$0.5\% uncertainty and negligible bias \citep{Kluge2024}. Because of the assumption of a single $\beta$-profile, the eSASS source detection algorithm tends to split bright clusters into multiple sources \citep[][]{Seppi2022A&A_eRASS1sim}. Using optical data, we address low significance detections related to the same halo and remove 413 sources sharing more than 70$\%$ of the galaxy members with another cluster whose extent likelihood is larger \citep[see][for more details]{Kluge2024}.

Given our purpose to constrain HOD models with eRASS1 clusters and groups, we build five volume-limited samples. Such a selection provides a representative subsample of the LSS compared to flux-limited samples \citep[see discussion in][]{Seppi2022A&A_eRASS1sim}. We build such samples by considering a reference flux of $F_{\rm X,lim}=4\times 10^{-14}\,{\rm erg/s/cm}^2$. Because of the varying sky coverage, the flux limit of the primary eRASS1 sample depends on the sky position. Nonetheless, this value provides an estimate of the average survey limit \citep[see][for details]{Bulbul2024}{}{}. We choose five redshift upper limits of 0.2, 0.3, 0.4, 0.5, and 0.6. We compute the luminosity threshold relative to the flux limit by $L_{X} = F_{\rm X,lim}\times 4\pi D_{L}^2(z)$, where $D_L(z)$ is the luminosity distance at the redshift, z \citep[][]{Hogg1999astro.ph..5116H}. We discard clusters located further away than each redshift threshold and fainter than the corresponding X-ray luminosity limit. The selection of volume-limited samples is shown in Fig. \ref{fig:Lx_z_volume_limited}. Our selection scheme basically provides a selection in count rate, which introduces a larger fraction of low luminosity groups in the low redshift samples compared to more luminous clusters at high redshift (see Sect. \ref{subsubsec:HOD}). This scheme allows us to study the evolution of HOD properties with X-ray luminosity and redshift. We obtain respectively 1650, 2993, 3732, 3670, and 3333 sources in the five samples. They span different cosmological volumes, from 0.68 to 15.16 Gpc$^3$. The volume is computed as the product of the comoving volume enclosed by the redshift limits and the survey area fraction. The number density varies from 2.2 to $24.45 \times 10^{-7}\,{\rm Mpc}^{-3}$. Table \ref{tab:z_sample} summarises the sample selection and their properties.

Motivated by the high purity of the catalogue thanks to $P_{\rm cont}$, we additionally focus on a flux-limited sample up to $z < 0.6$ to infer cosmological parameters. We apply a luminosity cut at 10$^{42.7}$ erg/s, the same as for the volume-limited S0 sample. This allows us to discard the faintest structures. This final sample consists of 6493 clusters, with an average redshift of $z=0.307$. This is the subsample covering the largest volume of $15.16\,{\rm Gpc}^3$ and is well suited for cosmological studies of the LSS of the Universe. 


\begin{table*}[]
    \centering
    \caption{Properties of the volume-limited eRASS1 cluster samples for our clustering study. We consider a flux limit of $F_{X,lim}=4\times10^{-14}$ erg/s/cm$^2$. We focus on a secure cluster sample with purity > 95$\%$ in the redshift range 0.1<z<0.6.}
    \label{tab:z_sample}
    \begin{tabular}{|c c c c c c c c c|}
        \hline
        \hline
        \rule{0pt}{2.2ex}
        \textbf{Label} &
        \multicolumn{3}{c}{\textbf{Selection}} & \textbf{<z>} & \textbf{N} & \textbf{Volume} [Gpc$^3$] & \textbf{Density} [10$^{-7}$Mpc$^{-3}$] & \textbf{S/N}\\
        & \textbf{z$_{\rm min}$} & \textbf{z$_{\rm max}$} & \textbf{L$_{\rm X,min}$} [erg/s/cm$^2$] & & & & & \\
        \hline
        \rule{0pt}{2.0ex}S0 & 0.1 & 0.2 & 42.7 & 0.153 & 1650 & 0.68 & 24.44 & 29.9\\ 
        S1 & 0.1 & 0.3 & 43.1 & 0.206 & 2993 & 2.33 & 12.85 & 39.4\\
        S2 & 0.1 & 0.4 & 43.4 & 0.264 & 3732 & 5.22 & 7.15 & 41.1\\
        S3 & 0.1 & 0.5 & 43.6 & 0.315 & 3670 & 9.49 & 3.87 & 35.4\\
        S4 & 0.1 & 0.6 & 43.8 & 0.369 & 3333 & 15.16 & 2.19 & 29.7\\
        \hline
        \hline
    \end{tabular}\\
    \footnotesize{\textbf{Notes}. The columns report the label, redshift and luminosity selection of each sample, the average redshift, the total number of sources, the cosmological volume, the source number density, and the signal-to-noise ratio of the clustering measurement explained in Sect. \ref{sec:method}.}
\end{table*}

\begin{figure}
    \centering
    \includegraphics[width=\columnwidth]{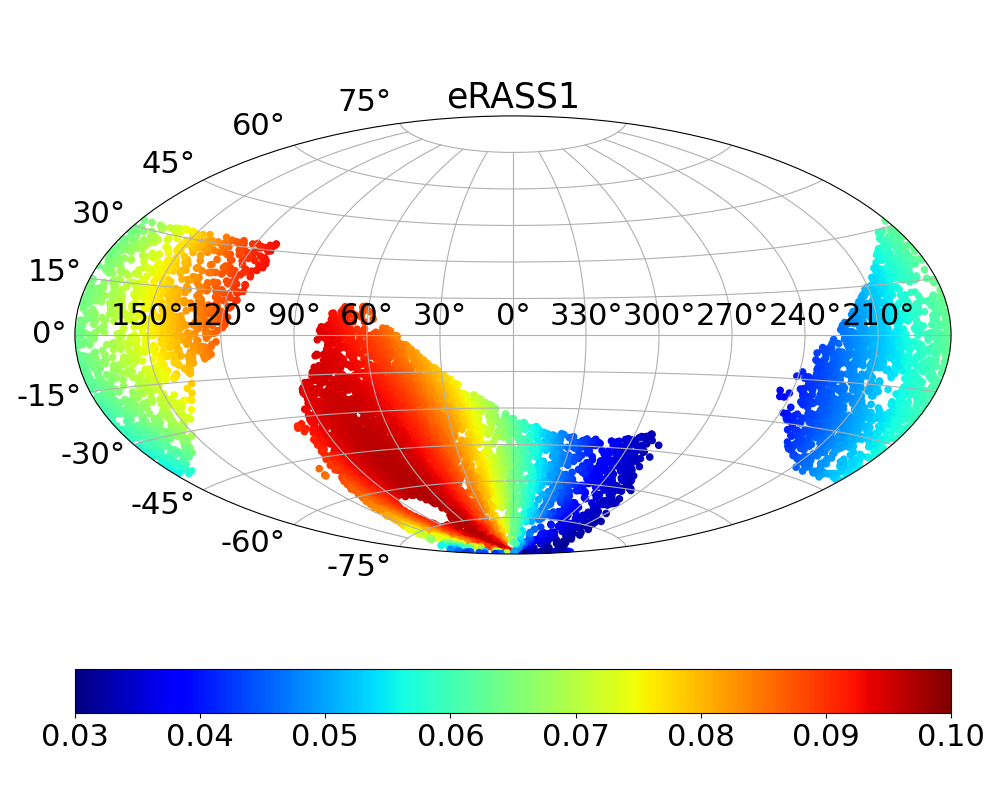}
    \includegraphics[width=\columnwidth]{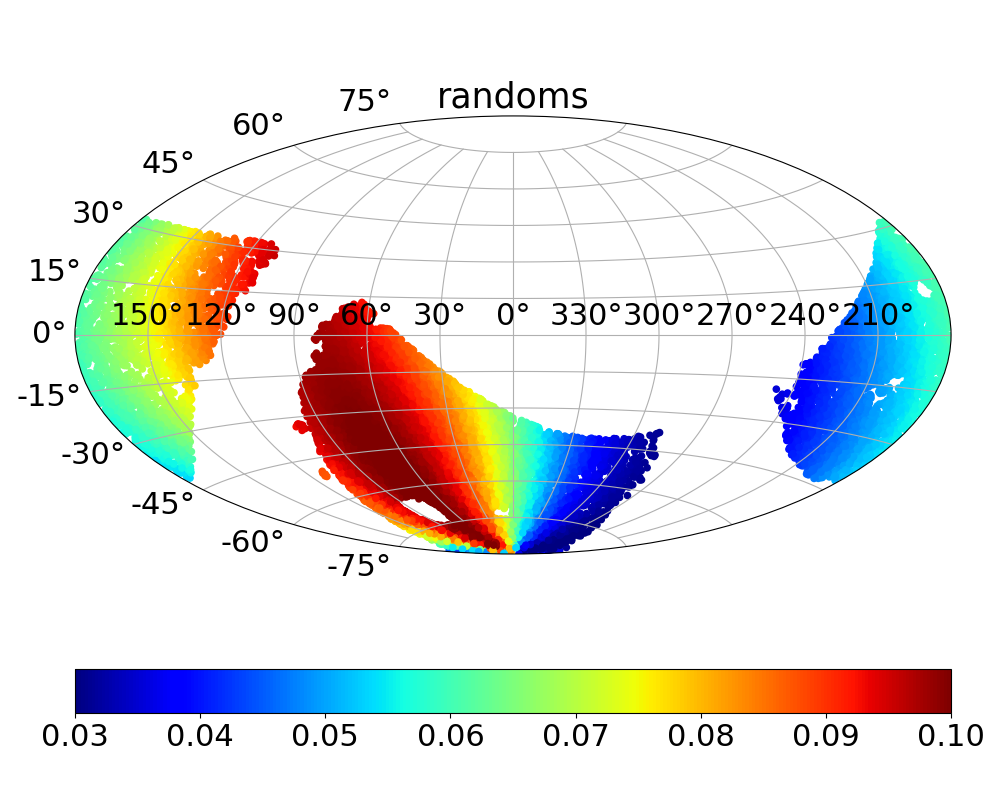}
    \caption{Angular distribution of the selected eRASS:1 clusters (top panel) and the random points (bottom panel), after the application of the masks. The points are color coded according to their normalized local density. The density of the random catalogue changes through the sky very similarly to the eRASS1 data.}
    \label{fig:sky_plot}
\end{figure}

\begin{figure}
    \centering
    \includegraphics[width=\columnwidth]{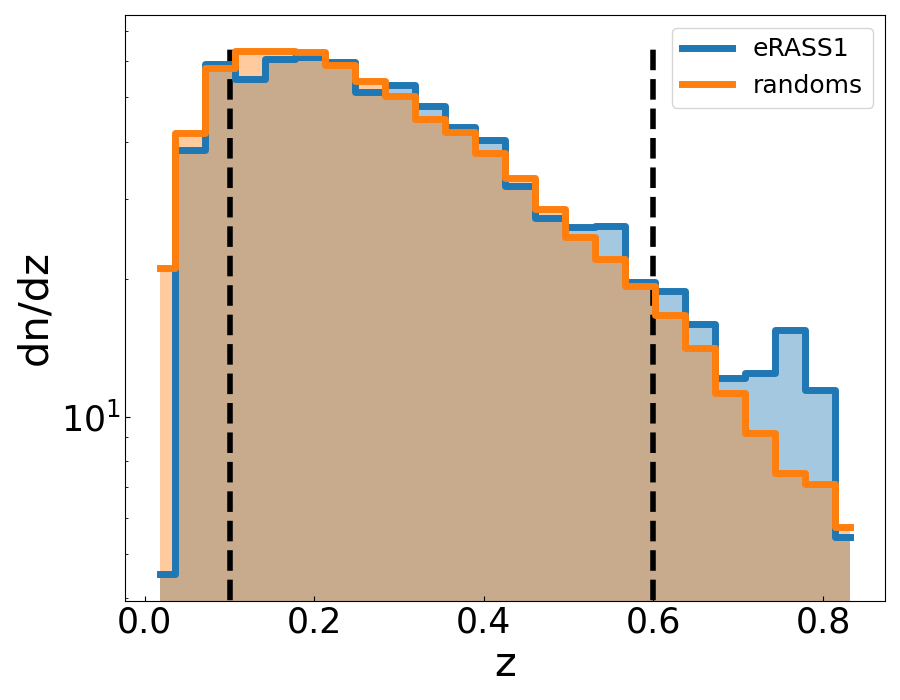}
    \caption{Probability density redshift distribution of the selected eRASS:1 clusters and the random points. The panel shows the total number of clusters (randoms) in each redshift bin divided by the size of the bin in blue (orange), normalized by the integral over the redshift range. Two black vertical lines denote the redshift range used in our analysis, between 0.1 and 0.6. The bump at redshift about 0.8 is due to a filter transition in the measurement of photo-z \citep[see][]{Kluge2024}. 
    }
    \label{fig:dndz}
\end{figure}

\subsection{Random catalog}
\label{subsec:random_cat}

The computation of the two-point correlation function (see Sect. \ref{sec:method}) requires the generation of random points uniformly distributed on the sky, following the same geometrical properties and the redshift distribution of the eRASS:1 cluster catalogue. We start from the random catalogues provided by the Legacy Survey DR10\footnote{\url{https://www.legacysurvey.org/dr10/files/\#random-catalogs-randoms}}. They have a spatial density of 50\,000 deg$^{-2}$. Following the optical selection of the eRASS1 primary sample \citep{Bulbul2024, Kluge2024}, we require coverage in the g, r, and z bands. 
We account for the cluster selection due to the varying eRASS1 exposure using the selection function from \citet{Clerc2023}. It depends on count rate, redshift, exposure time, background and absorption. We randomly assign count rates and redshifts to the random points by drawing samples from a Gaussian kernel distribution\footnote{\texttt{scipy,} \url{https://scipy.org/}} \citep[][]{Virtanen2020SciPy-NMeth}, generated from the eRASS1 digital twin from \citet[][]{Seppi2022A&A_eRASS1sim}. The simulation is based on the cluster model from \citet[][]{Comparat2020Xray_simulation}. We compute the probability of detection for each random point after fixing the exposure and background values from the eRASS:1 maps, and the nH values from \citet{HI4PI2016A&A...594A.116H} at the position of each point. If the detection probability extracted from the selection function is larger than a random number between 0 and 1, we keep the random point in the catalogue. 

This method provides a random catalogue that closely follows the angular and redshift distribution of the eRASS1 clusters by construction. The incompleteness of the eRASS1 sample, encoded in the selection function itself, is also naturally accounted for by our strategy. We show the spatial distribution on the sky in Fig. \ref{fig:sky_plot}. The top panel shows the eRASS1 clusters, and the bottom one the random points. They are color-coded according to their local density, estimated by a local kernel density estimation with \texttt{scipy}. 
After the selection function cut, the random catalogues are about 50 times larger than the eRASS1 samples. 

The number density as a function of redshift of the eRASS1 clusters and the random points is shown in Fig. \ref{fig:dndz}. The histograms for the random populations are normalized to match the highest point of the cluster ones. The panel shows the two samples, respectively in blue and orange. The shape of the randoms redshift distribution captures the trend in the real data. We notice how the redshift distribution of the real eRASS1 is more peaked around $z=0.8$, due to an optical filter transition. Although this is not a problem for the main cluster count analysis \citep{Ghirardini2024} because of the redshift uncertainty evaluation, we restrict our analysis in the $0.1<z<0.6$ range, where the redshift distributions of the random is similar to the data. We tested the reliability of our randoms by measuring the angular clustering of the eRASS1 clusters as a function of angular separation $\theta$. We obtain a correlation function scaling with $\theta^{-0.82}$, in agreement with the expected power law scaling with a slope of -0.8 \citep[see e.g.,][]{Baugh1993MNRAS_wtheta, Maller2005ApJ_wtheta}.

We also tried a more traditional method following the sensitivity maps \citep[][]{Georgakakis2008MNRAS.388.1205G}, used to construct random catalogues of point sources \citep[][]{Comparat2023A&Aagnclustering}. Although such a method is excellent for point sources, it is not ideal for extended ones. Within eSASS, the \texttt{ersensmap} task computing sensitivity maps for extended sources assumes a simple $\beta$-profile. Different works showed that the cluster detection is more complex \citep[][]{Clerc2018A&A...617A..92C, Seppi2022A&A_eRASS1sim}. The random catalogue generated with this method may therefore introduce spurious clustering in the measurement of the correlation function. They are not used here.


\section{Method}
\label{sec:method}

We measure the redshift-space two-point auto-correlation function for eRASS1 clusters on a two-dimensional grid of separation perpendicular and parallel to the line sight, respectively denoted as $r_{\rm p}$ and $\pi$ \citep{Fisher1994MNRAS_IRASclustering}. 


This strategy allows separating spatial correlations from redshift-space distortions due to large-scale bulk motions and intrinsic velocities \citep[RSD,][]{Kaiser1987MNRAS.227....1K}. We use the Landy-Szalay estimator \citep{Landy1993ApJ_2pcfestimator}:
\begin{equation}
    \xi(r_{\rm p},\pi) = 1 + \frac{N_{RR}}{N_{DD}}\frac{DD(r_{\rm p},\pi)}{RR(r_{\rm p},\pi)} - 2 \frac{N_{RR}}{N_{DR}}\frac{DR(r_{\rm p},\pi)}{RR(r_{\rm p},\pi)},
    \label{eq:LS93}
\end{equation}
where $DD(r_{\rm p},\pi)$, $RR(r_{\rm p},\pi)$, and $DR(r_{\rm p},\pi)$ are the number of pairs in two-dimensional bins of $r_{\rm p}$ and $\pi$ for couples of data, randoms, and data-random. The total pairs of data, randoms, and data-randoms are respectively denoted by $N_{DD}$ = $N_D$($N_D$-1)/2, $N_{RR}$ = $N_R$($N_R$-1)/2, and $N_{DR}$ = $N_DN_R$. This estimator minimizes variance and finite volume effects \citep[][]{Hamilton1993ApJ_clustering}. \\
We minimize the impact of RSD by computing the correlation function projected along the line of sight:
\begin{equation}
    w_{\rm p}(r_{\rm p}) = 2\int_0^{\pi_{\rm max}} \xi(r_{\rm p},\pi)d\pi.
    \label{eq:wprp}
\end{equation}
For our study, we consider ten logarithmic bins of separation r$_p$ between 1 and $80\,{\rm Mpc}/h$. 
We choose $1\,{\rm Mpc}/h$ bins for $\pi$.
In principle, $\pi_{\rm max}$ should extend to infinity. In practice, we use $\pi_{\rm max} = 100\,{\rm Mpc}/h$ in this study. We tested other values of $\pi_{\rm max}$ equal to $80$ and $120\,{\rm Mpc}/h$ and found that they do not increase the signal-to-noise ratio compared to the chosen limit of $100\,{\rm Mpc}/h$. 
For our purpose, this limit ensures the inclusion of correlated pairs and minimizes the noisy contribution of distant ones. It allows including pairs whose separation is well within the typical uncertainty on the photometric redshift with a mean scatter of $0.005\times(1+z)$ (about 25 comoving Mpc at $z=0.3$), see \citep{Kluge2024}. 
We compute pair counts and the correlation function in Eqs. \ref{eq:LS93} and \ref{eq:wprp} with the software \texttt{Corrfunc} \footnote{\url{https://corrfunc.readthedocs.io/}} \citep{Sinha19_corrfunc, Sinha2020MNRAS_corrfunc}. 

\begin{figure}
    \centering
    \includegraphics[width=\columnwidth]{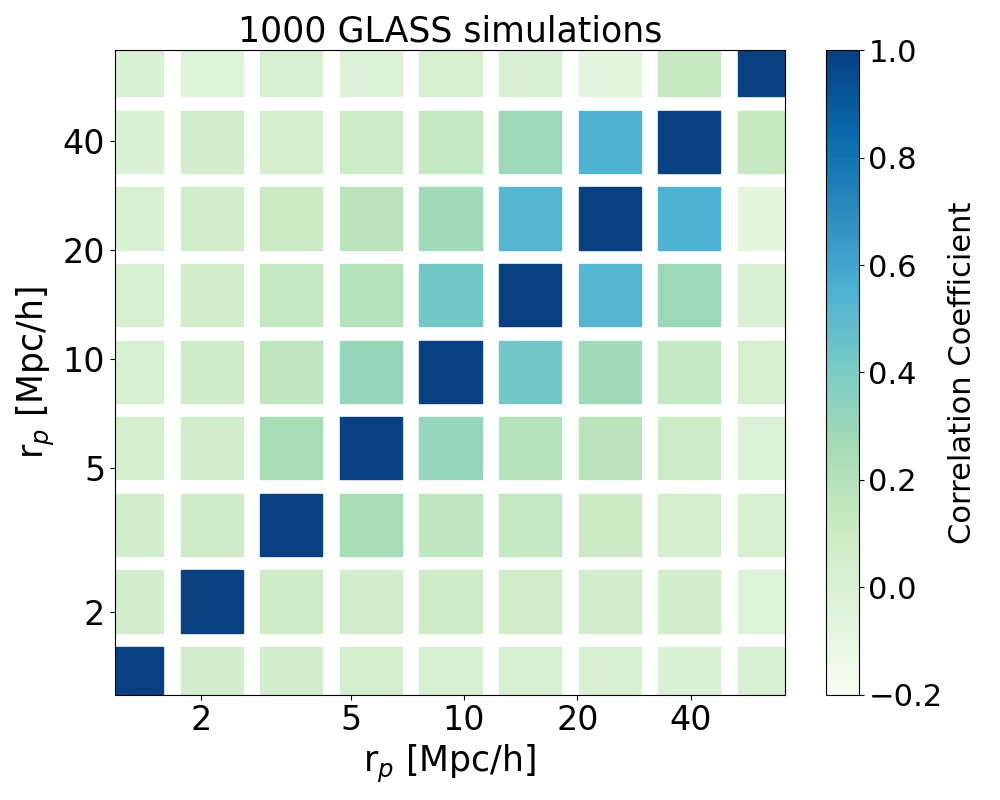}
    \caption{Jackknife correlation matrix $C_{i,j}$/$\sqrt{C_{i,i}C_{j,j}}$ for the projected correlation function of the eRASS1 cluster sample described in Sect. \ref{sec:data}. The covariance matrix for the different samples presented in Sect. \ref{subsec:erosita_data} is computed via jackknife (see Eq. \ref{eq:covmat_combined}.) 
    }
    \label{fig:correlation_matrix}
\end{figure}

We estimate the diagonal terms of the covariance matrix between different bins of separation with a jackknife method on the eRASS1 data. 
We create multiple subsamples of the cluster and random catalogues by removing different areas of about 54 deg$^2$ each time. Every area corresponds to a single HEALPix tile with NSIDE~=~8\footnote{\url{https://healpix.sourceforge.io/credits.php}}. Such a pixelization divides the full sky into 768 pixels. We remove one area, compute a correlation function $w_{p}^k$ and iterate the process. Given the area covered by eRASS1, the removal of one pixel reduces the area by about 0.43$\%$. 
We compute the jackknife covariance matrix according to 
%
\begin{equation}
    \hat{\textbf{C}_{\rm i,j}} = \frac{N_{\rm S}-1}{N_{\rm S}} \sum_{\rm k = 1}^{\rm N_{\rm S}} (w_{p,i}^k - \overline{w_{\rm p,i}})(w_{p,j}^k - \overline{w_{\rm p,j}}),
    \label{eq:cov_mat}
\end{equation}
where $w_{\rm p,j}^k$ is the value of the k correlation function at separation j, $\overline{w_{\rm p,i}}$ is the average value of the correlation function at separation i, and $\text{N}_{\rm S}$ is the number of subsamples \citep[][]{Norberg2009MNRAS.396...19N}. 
\\
To estimate the crossed terms of the covariance matrix we generate 1000 simulations using the Generator for Large-Scale Structure \citep[\texttt{GLASS},][]{Tessore2023_GLASS}, creating shells of 200 comoving Mpc with a source density of 1 deg$^{-2}$. With this setup, we can study the correlation between separations with a source sky density that is similar to the average for eRASS1. We measure the correlation function for each realization, compute a covariance matrix according to Eq. \ref{eq:cov_mat}, and derive the correlation matrix $c_{GLASS i,j}$. Because these simulations do not include the selection function and the geometry of the survey, we only use them to estimate the relative correlation between crossed terms of the covariance matrix. We rely on the jackknife method for the diagonal terms. We combine them to generate the final covariance matrix:
\begin{equation}
    C_{\rm i,j} = c_{\rm GLASS i,j} \times \big[ \hat{C_{\rm i,i}} \otimes \hat{C_{\rm j,j}} \big]. 
    \label{eq:covmat_combined}
\end{equation}
The correlation matrix $C_{i,j}$/$\sqrt{C_{i,i}C_{j,j}}$ is shown in Fig. \ref{fig:correlation_matrix}.
The panel shows the correlation matrix $c_{\rm GLASS\ i,j}$. The latter is rescaled by the diagonal terms for each sample according to Eq. \ref{eq:covmat_combined}. The figure highlights the low correlation between small and large scales, due to the separation between the 1-halo and 2-halo terms, around $2\,{\rm Mpc}/h$.

Given the excellent quality of the photometric redshifts, we do not account for additional redshift uncertainties \citep[see][]{Kluge2024}. 
We further test this with the eRASS1 mock from \citet[][]{Seppi2022A&A_eRASS1sim}. As explained in the previous paragraph, we compute the projected correlation function for the volume-limited sample spanning $0.1<z<0.6$. At first, we use the spectroscopic redshifts in the simulation for distance conversions and measure the true $w_{\rm p, T}$. Secondly, we add a normal scatter of $F \times (1+z)$ to the spec-$z$, with three values of the factor F equal to 0.005, which is the eRASS1 case, but also worse-case scenarios of 0.008 and 0.01. We measure a $w_{\rm p, F}$ for each one. We compute the average shift of the correlation function due to the redshift error as 
\begin{equation}
E = \frac{w_{\rm p,T} - w_{\rm p,F}}{w_{\rm p,T}}.
\label{eq:zerr_wprp}
\end{equation}
We obtain shifts of E equal to 1.8$\%$, 12.9$\%$, and 15.3$\%$. The covariance matrix gives a relative uncertainty of 33$\%$. We conclude that redshift errors have a minimal impact on our measurement, especially for the eRASS1 case with 0.5$\%$ scatter.

\subsection{Models}
\label{subsec:Models}

The correlation function is the real-space analogous to the matter power spectrum:
\begin{equation}
    \xi(r,z) = \frac{1}{2\pi^2} \int dk k^2 P_{\rm M}(k,z=0)D^2(z)\frac{\sin(kr)}{kr},
    \label{eq:corr_func}
\end{equation}
where $P_{\rm M}(k,z)$ is the matter power spectrum and $D^2(z)$ is the growth factor normalized to 1 at z=0. We denote physical distances with $r$, and redshift-inferred distances as $s$.\\
For the modelling, we convert $\pi$ and r$_{\rm p}$ to the redshift-space radial distance and the angle between the line of sight and the direction of each pair according to:
\begin{align}
    s =& \sqrt{\pi^2 + r_{\rm p}^2} \nonumber \\
    \mu =& \frac{\pi}{s}.
    \label{eq:s_mu}
\end{align}
We model the correlation function in redshift space according to
\begin{align}
    \xi(s,\mu) =&\ \xi_0(s)P_0(\mu) + \xi_2(s)P_2(\mu) + \xi_4(s)P_4(\mu)  \nonumber \\
    \xi_0(s) =&\ \Big[1+\frac{2}{3}\beta + \frac{1}{5}\beta^2\Big]\xi(r) \nonumber \\
    \xi_2(s) =&\ \Big[\frac{4}{3}\beta + \frac{4}{7}\beta^2\Big](\xi(r) - \overline{\xi}(r)) \nonumber \\
    \xi_4(s) =&\ \frac{8}{35}\beta^2 \Big[ \xi(r) + \frac{5}{2}\overline{\xi}(r) - \frac{7}{2}\overline{\overline{\xi}}(r)),
    \label{eq:2pcf_redshiftspace}    
\end{align}
where $P_\ell$ are Legendre polynomials, $\xi_0$, $\xi_2$, $\xi_4$ are the monopole, dipole, and quadrupole moments of the 3D correlation function, $\overline{\xi}(r)$ and $\overline{\overline{\xi}}(r)$ are integrals of the real space correlation function \citep[see][]{Hamilton1992ApJ_multipoles}, and $\beta = f/b$ is the ratio between the cosmic growth rate and the bias \citep{Hamilton1992ApJ_multipoles, Valageas2012A&A...547A.100V, Jeong2015MNRAS_redshiftspace_2pcf}. The growth rate is defined as the derivative of the growth factor $D$ with respect to the scale factor $a$: $f = d\ln D / d\ln a$ \citep{Peacock2001Natur.410..169P}.
We finally integrate the model in Eq. \ref{eq:2pcf_redshiftspace} according to Eq. \ref{eq:wprp} to obtain a prediction of the two-point correlation function projected along the line of sight.

\subsubsection{HOD}
\label{subsubsec:HOD}

\begin{table*}[]
    \caption{Priors and posteriors for the HOD and the derived parameters on the 5 volume-limited eRASS1 samples (see Table \ref{tab:z_sample} and Fig.\ref{fig:Lx_z_volume_limited}).}
    \label{tab:parameter_HOD}
    \centering
    \begin{tabular}{|c | c | c | c | c | c | c |}
    \hline
    \hline
    \multirow{2}*{Parameter} & \multirow{2}*{Prior} & \multicolumn{5}{c|}{Posterior}\\
    & & S0 & S1 & S2 & S3 & S4 \\
    \hline
        \rule{0pt}{2.2ex}
        $M_{\rm min}$ & $\mathcal{U}(13, 16)$ & 14.73$^{+0.93}_{-0.35}$ & 14.90$^{+0.87}_{-0.24}$ & 14.94$^{+0.84}_{-0.26}$ &
        15.02$^{+0.63}_{-0.63}$ &
        15.04$^{+0.61}_{-0.72}$\\
        \rule{0pt}{2.2ex}
        $\sigma_{\log M}$ & $\mathcal{U}(0.01, 2.0)$ & 0.82$^{+0.49}_{-0.46}$ & 0.66$^{+0.50}_{-0.34}$ &  0.62$^{+0.46}_{-0.32}$ &  0.58$^{+0.23}_{-0.40}$ & 0.57$^{+0.20}_{-0.37}$\\
        \rule{0pt}{2.2ex}
        $\alpha_{\rm sat}$ & $\mathcal{U}(0.1, 3)$ &   1.25$^{+0.89}_{-0.54}$ &   1.48$^{+0.85}_{-0.60}$ & 1.41$^{+0.82}_{-0.58}$ &
        1.65$^{+1.04}_{-0.93}$ & 1.67$^{+1.05}_{-1.08}$\\
        \rule{0pt}{2.2ex}
        $M_{\rm sat}$ & $\mathcal{U}(13.0, 16.5)$ & 15.32$^{+0.34}_{-0.33}$ &  15.70$^{+0.30}_{-0.27}$ & 15.63$^{+0.29}_{-0.28}$ & 15.79$^{+0.36}_{-0.30}$ & 15.76$^{+0.42}_{-0.35}$ \\
        \rule{0pt}{2.2ex}
        $\alpha_{inc}$ & fixed & 0.55 & 0.55 & 0.55 & 0.55 & 0.55 \\
        \rule{0pt}{2.2ex}
        $M_{inc}$ & fixed & 14.87 & 14.95 & 15.06 & 15.16 & 15.25 \\
        \hline
        \rule{0pt}{2.0ex}
        $f_{sat}$ & - & <14.9\% & <9.3\% & <9.9\% & <6.8\% & <8.1\% \\
        \rule{0pt}{2.0ex}
        b & - & 2.95 $\pm$ 0.21 & 3.35 $\pm$ 0.23 & 3.69 $\pm$ 0.27 & 4.15 $\pm$ 0.42 & 4.34 $\pm$ 0.62 \\
        \rule{0pt}{2.0ex}
        $\overline{M_{\rm halo}}$ [10$^{14}$ M$_\odot$] & - & 3.09 $\pm$ 0.48 & 3.54 $\pm$ 0.51 & 3.83 $\pm$ 0.55 & 4.44 $\pm$ 0.89 & 4.38 $\pm$ 1.11 \\
        \rule{0pt}{2.0ex}
        $\overline{L_{\rm X}}$ [10$^{43}$ erg/s] & - & 4.63$\pm$0.97 & 7.38$\pm$0.98 & 10.21$\pm$0.98 & 13.68$\pm$0.99 & 17.76$\pm$0.99 \\
    \hline
    \end{tabular}
    \tablefoot{$\mathcal{U}({\rm min}, {\rm max})$ indicates a uniform prior between min and max values. The parameters with fixed prior have been estimated using the eRASS1 digital twin \citep[][]{Seppi2022A&A_eRASS1sim}. The bottom part of the table shows derived parameters: the fraction of satellites $f_{sat}$ (see Eq. \ref{eq:frac_sat}), the large-scale halo bias, and the average halo mass. The average X-ray luminosity for each sample is reported in the last row. The upper limits extend to the 84th percentile.}
\end{table*}

We interpret the eRASS1 correlation function using the Halo Occupation Distribution formalism \citep[][]{Kravtsov2004ApJHOD, Zheng2005ApJ_HOD}. 
Following the example of \cite{More2015ApJ_HOD_aum}, we parametrize the HOD models as 
\begin{align}
    &N_{\rm C}(M | M_{\rm min}, \sigma_{\rm \log M}) = f_{\rm inc}(M) \frac{1}{2} \Big[1+\text{erf} \Big( \frac{\log(M/10^{M_{\rm min}})}{\sigma_{\rm \log{M}}}\Big) \Big] \notag \\
    &N_{\rm S}(M | M_{\rm sat}, \alpha_{\rm sat}) = N_{\rm C}\Big(\frac{M - 10^{\rm M_{\rm sat}-1}}{10^{\rm M_{\rm sat}}} \Big)^{\rm \alpha_{\rm sat}} \notag \\
    &f_{\rm inc}(M | \alpha_{\rm inc}, M_{\rm inc}) = \text{max}[0, \text{min}[1, 1+\alpha_{\rm inc}(\log M - \log M_{\rm inc})]].
    \label{eq:HOD}
\end{align}
Our model depends on four free parameters: $M_{\rm min}$ is the typical mass of a dark matter halo that starts being populated by sources in our sample, $\sigma_{\log{M}}$ describes the sharpness of the transition between unpopulated and populated dark matter haloes, $M_{\rm sat}$ and $\alpha_{\rm sat}$ define the power law describing the average number of satellites, i.e. low mass haloes companion to massive clusters. We also account for the incompleteness in the eRASS1 cluster samples, based on the eRASS1 digital twin \citep[][]{Seppi2022A&A_eRASS1sim} and fit it with the incompleteness function in Eq. \ref{eq:HOD} using the \texttt{curve\_fit} package \footnote{\url{https://scipy.org/}}. We do not find a significant dependence on the slope $\alpha_{\rm inc}$ for different samples. Therefore, we fix it to 0.55. We report the best-fit values in Table \ref{tab:parameter_HOD}. We find a progressively lower $M_{inc}$ parameter, decreasing from 15.25 for the S4 sample to 14.87 for the S0 one. It means that deeper samples start being incomplete at lower masses. \\ 
Finally, we infer the fraction of satellites by weighting the number of satellites by the halo mass function compared to the sum of satellites and centrals:
\begin{equation}
    f_{\rm sat} = \frac{\int\int N_{\rm S}(M)\dfrac{dn}{dM}dVdM}{\int \int N_{\rm TOT}(M) \dfrac{dn}{dM}dVdM},
    \label{eq:frac_sat}
\end{equation}
where $N_{\rm TOT} = N_{\rm C} + N_{\rm S}$ and $dn/dM(M)$ is the dark matter halo mass function. We use the halo mass function model from \citet[][]{Tinker2008}. We tested other halo mass functions from \citet[][]{Despali2016} and \citet[][]{Seppi2021A&A...652A.155S}. We verified that the chosen mass function model does not significantly affect our results.

We model the power spectrum according to \citet[][]{Asgari2023arXivhalo_model_review}. The power spectrum prediction includes multiple terms as follows: 
\begin{align}
    P(k,z) = & 2P_{c,s}^{1h}(k,z) + P_{s,s}^{1h}(k,z) + P_{c,s}^{1h}(k,z) + \nonumber \\
    & P_{c,c}^{2h}(k,z) + 2P_{c,s}^{2h}(k,z) + P_{s,s}^{2h}(k,z).  
    \label{eq:pk_halo}
\end{align}
The subscripts $c$ and $s$ refer to central and satellite objects. The overscripts $1h$ and $2h$ indicate 1-halo and 2-halo terms.  The correlation between central objects in two separate haloes is then $P_{c,c}^{2h}$, the one between satellites in the same halo is $P_{s,s}^{1h}$. Each term is a mass integral including the quantities in Eq. \ref{eq:HOD}. The large-scale halo bias is finally inferred by convolving the \citet{Tinker2010ApJ_bias} model with central, satellites, and the halo mass function in a mass integral \citep{Berlind2002ApJHOD}, providing more flexibility compared to a strict choice of bias model thanks to the self-consistent HOD modelling. We refer the reader to \citet[][]{More2015ApJ_HOD_aum} for more details on the formalism for each term of Eq. \ref{eq:pk_halo} and to \citet{vandenBosch2013MNRAS_HODmodeling_AUM} for the general theoretical framework.

We predict HOD models of the correlation functions with \texttt{AUM}\footnote{\url{https://github.com/surhudm/aum}} \citep[][]{vandenBosch2013MNRAS_HODmodeling_AUM, More2021ascl.soft08002M_AUM}. This software uses the $M_{200{\rm b}}$ mass definition, corresponding to the mass enclosed by a radius encompassing an average density equal to 200 times the background matter density. Therefore, we will quote HOD results in terms of $M_{200{\rm b}}$ in the rest of the article, if not otherwise stated.

\subsubsection{Cosmology}
\label{subsubsec:Cosmo}

For the cosmological study, we focus on a flux-limited sample of 6493 clusters, spanning the redshift range 0.1~--~0.6, with X-ray luminosity larger than $5\times 10^{42}\,{\rm erg/s}$ (a cut in log at 42.7). This sample covers the largest cosmological volume of $15.16\,{\rm Gpc^3}$. The sample selection is explained in Sect. \ref{subsec:erosita_data}. 

A key ingredient of a clustering analysis is the prescription for the large-scale halo bias model. A common strategy consists of using a scaling relation approach to obtain halo masses from observables such as X-ray luminosity or richness \citep[][]{Lindholm2021A&A...646A...8L, Lesci2022A&A_clustering}. More recently, the development of emulators allows a fast prediction of clustering measurements that intrinsically include the bias \citep[][]{Nishimichi2019ApJ...884...29N, Sunayama2023arXiv230913025S_redmapperDR8}. Although a theoretical estimate of the bias in the cluster regime is possible \citep[][]{Sheth1999MNRAS_LSbias}, one limitation is the precision and accuracy of different halo bias models. Clusters of galaxies are strongly biased tracers of the LSS, and various models show discrepancies even >20$\%$ in this regime \citep[][]{Tinker2010ApJ_bias, Bhattacharya2011, Comparat2017, Lindholm2021A&A...646A...8L}. The choice of a specific model at fixed mass may bias cosmological results. Therefore, we take advantage of our HOD approach to model the bias in a self-consistent way, with no prior assumption on halo mass as explained in Sect. \ref{subsubsec:HOD}. We develop a cosmological fitting method that models the large-scale halo bias together with halo occupation in a flexible manner. We use $\sigma_{\log M}$, $\alpha_{\rm sat}$, and $M_{\rm sat}$ as nuisance parameters, sampling a uniform prior containing the best-fit HOD results of each volume-limited sample in Table \ref{tab:z_sample}. With this method, we are not affected by fixed parameters obtained from the HOD fits in a fixed cosmology. We sample $M_{\rm min}$ together with cosmological parameters to account for the cosmological dependency of the halo bias on the clustering amplitude.

We predict models of the correlation function with the AUM software described in the previous section. It predicts power spectra based on \citet[][]{EisensteinHu1998ApJ...496..605E} for the linear power spectrum and the transfer function $T^2(k,z)$, and \citet[][]{Smith2003MNRAS.341.1311S} for the non linear regime. 
We use the same binning scheme as for the HOD modelling. The impact of nonlinearities \citep[][]{Angulo2021MNRAS_BACCO, Angulo2022_reviewSIMS} is intrinsically included in the HOD approach. Extending this study to even larger scales $\gtrsim 100\,{\rm Mpc}/h$ requires additional modelling of the baryonic acoustic oscillations (BAO) peak \citep[][]{Veropalumbo2016MNRAS.458.1909V_BAO}. The moments of $\xi(s)$ in Eq. \ref{eq:2pcf_redshiftspace} are computed by convolving the non-linear matter power spectrum with Bessel functions using \texttt{FFTLog} \citep[][]{Hamilton2000MNRAS_FFTLog} within \texttt{GSL}\footnote{\url{http://www.gnu.org/software/gsl}}.

We account for the \citet{Alcock1979Natur.281..358A} (AP) geometrical distortions of the correlation function via the $\alpha_\parallel$ and $\alpha_\perp$ parameters, defined according to 
\begin{align}
    \pi &= \alpha_\parallel \pi^{\rm fid} \nonumber \\
    r_{\rm p} &= \alpha_\perp r_{\rm p}^{\rm fid} \nonumber \\
    \alpha_\parallel &= \frac{H^{\rm fid}(z)}{H(z)}  \nonumber \\
    \alpha_\perp &= \frac{D_{\rm A}(z)}{D_{\rm A}^{\rm fid}(z)},
    \label{eq:alpha_perpendicular}
\end{align}
where $H(z)$ is the Hubble parameter, $D_{\rm A}(z)$ is the angular diameter distance, and the overscript 'fid' denotes the quantity computed by assuming a fiducial cosmology \citep[the one from][in this study]{Planck2020A&A...641A...6P} to convert redshifts to distances. The model of the correlation function is then evaluated at separation equal to $\alpha_\perp r_{\rm p}^{\rm fid}$ ($\alpha_\parallel \pi^{\rm fid}$) in the direction perpendicular (parallel) to the line of sight.

We follow the evolution mapping approach from \citet[][]{Sanchez2020PhRvD_littleh, Sanchez2022MNRAS_evomapping} to infer cosmological parameters in an $h$-independent way. We consider physical densities $\omega_i=\dfrac{8\pi G}{3H^2_{\rm 100}}\rho_i$, where $\rho_i$ is the density of each component at the present day, and $H_{\rm 100}=100\,{\rm km/s/Mpc}$. We fix the value of the scalar spectral index and the baryons density according to the results of \citet{Planck2020A&A...641A...6P} to $n_{\rm s}=0.9665$ and $\omega_{\rm b}=0.02242$. Given that clustering studies with galaxy clusters strongly depend on the matter density, we let the physical cold dark matter density $\omega_{\rm c}$ free to vary. We free parameters affecting the evolution of the power spectrum, such as its normalization $A_{\rm s}$, and the physical dark energy density $\omega_{\rm DE}$. We fix the dark energy equation of state to $w_0=-1$ and $w_a=0$, and assume a flat Universe $\Omega_K=0$. With this approach, the Hubble parameter is recovered by construction as $h=\sqrt{\omega_{\rm c} + \omega_{\rm b} + \omega_{\rm DE}}$.

\begin{figure}
    \centering
    \includegraphics[width=\columnwidth]{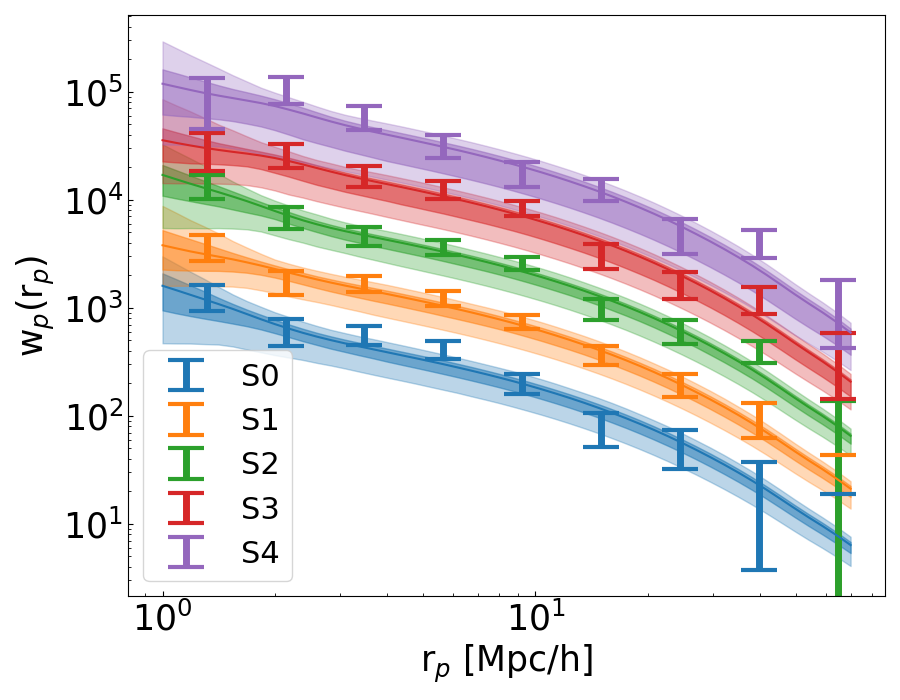}
    \caption{Projected correlation functions of the volume selected eRASS1 clusters samples. Each color denotes one sample. The error bars show the measurement, the solid line is the best fit model, and the shaded areas denote the 1-$\sigma$ and 2-$\sigma$ uncertainty on the model. For clarity, we shift the measurements and models by $e^{0,1,2,3,4}$ from S0 to S4.}
    \label{fig:HOD_wprp}
\end{figure}

\subsection{Likelihood}

Given the HOD parameterisation, we require our best fit to maximize a Gaussian likelihood. 
We compute it according to 
\begin{align}
    \log\mathcal{L} = \frac{1}{2}[\textbf{D}-\textbf{M}(\theta)]^T \textbf{C}_{ij}^{-1}[\textbf{D}-\textbf{M}(\theta)],
    \label{eq:likelihood}
\end{align}
where $D$ is the data array for each projected correlation function, $M$ is the $w_{\rm p}(r_{\rm p})$ model that depends on the array of parameters $\theta$. We fit five HOD models independently for each volume-limited sample in Table \ref{tab:z_sample}. Each sample has a different covariance matrix, as explained in Sect. \ref{sec:method}. We use the same likelihood form to fit cosmological parameters together with the HOD ones (see Sect. \ref{subsubsec:Cosmo}).\\
We derive posterior probability distributions and the Bayesian evidence with the nested sampling Monte Carlo algorithm MLFriends \citep[][]{Buchner2014, Buchner2019} using the \texttt{UltraNest}\footnote{\url{https://johannesbuchner.github.io/UltraNest/}} package \citep[][]{Buchner2021_ultranest}. We quote as best-fit parameters the 50$\%$ quantile of the marginalized posterior distributions. The uncertainty on the parameters includes the 16th-84th percentile. The upper limits extend to the 84th percentile.

\section{Results}
\label{sec:results}
We present our results in this section. The HOD interpretation is described in Sect. \ref{subsec:HOD_result}. 
The cosmological implications are reported in Sect. \ref{subsec:cosmo_results}.

\subsection{HOD interpretation}
\label{subsec:HOD_result}

\begin{figure}
    \centering
    \includegraphics[width=\columnwidth]{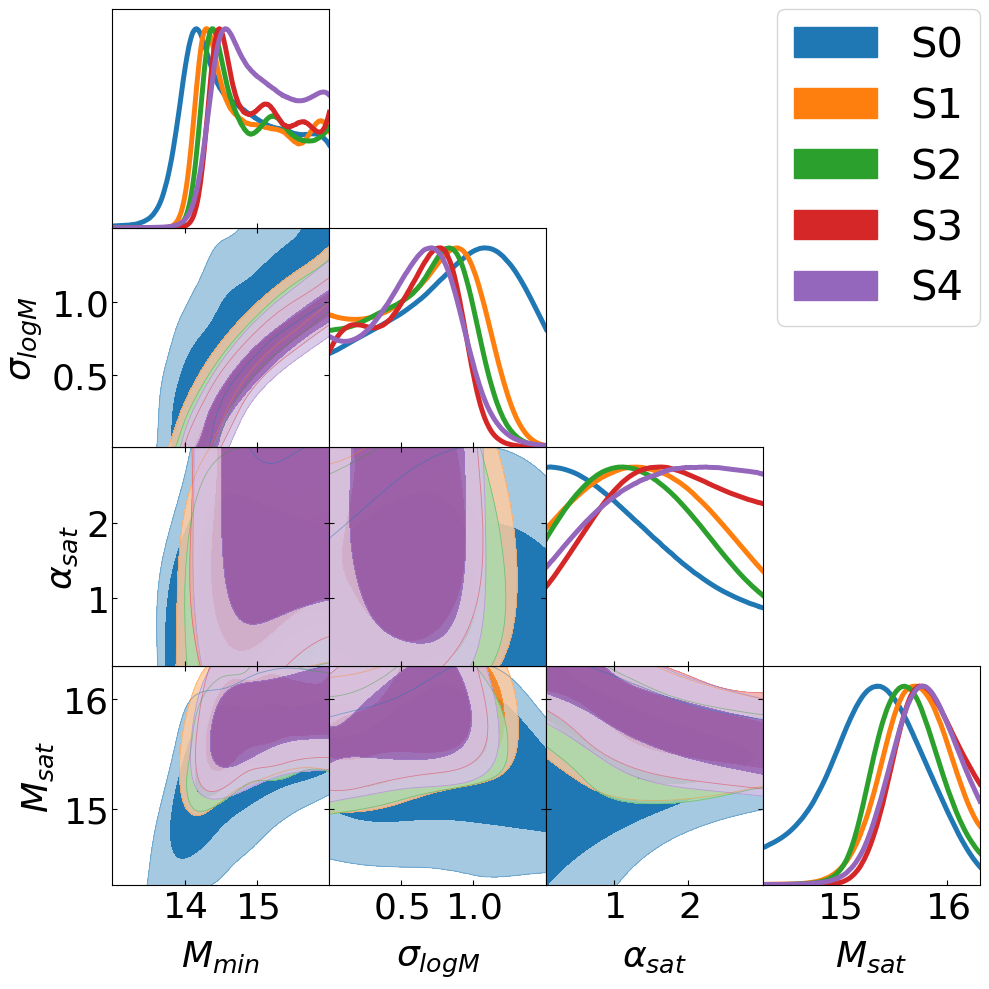}
    \caption{Marginalized posterior distributions of the best fit HOD parameters. The filled 2D contours show the 1-$\sigma$ and 2-$\sigma$ confidence levels of the posteriors after convolution with the uniform priors. The model is given by Eq. \ref{eq:HOD}. The corresponding one-dimensional parameter constraints are reported in Table \ref{tab:parameter_HOD}.
    }
    \label{fig:HOD_corner}
\end{figure}

\begin{figure}
    \centering
    \includegraphics[width=\columnwidth]{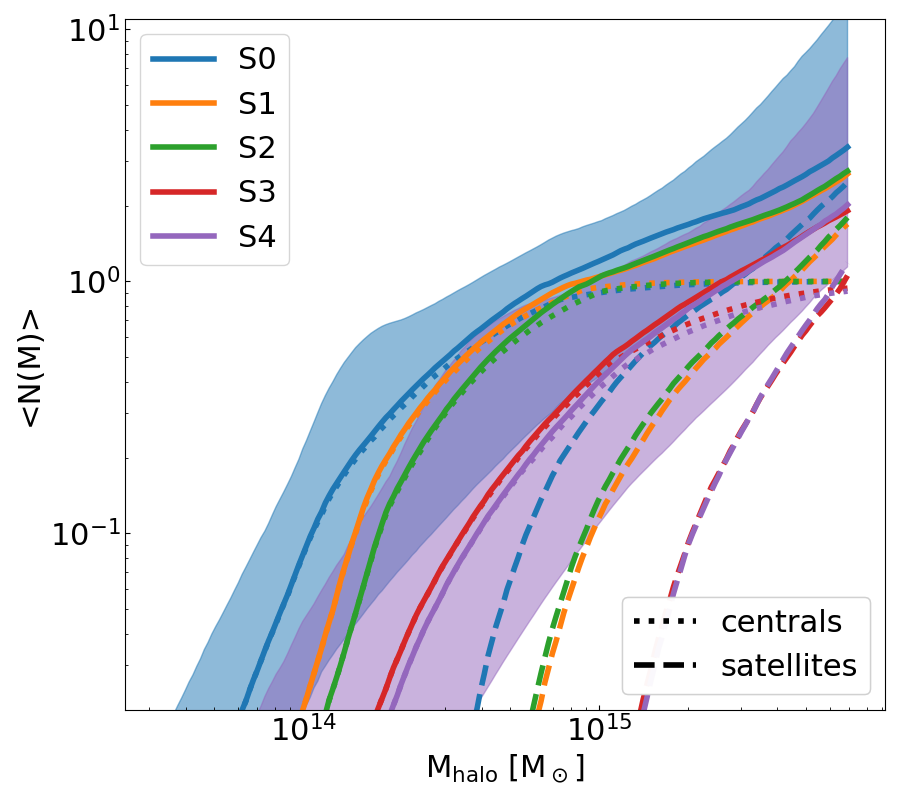}
    \caption{Derived distribution of the eRASS1 clusters and groups population in dark matter haloes (solid lines), divided in central (dashed) and satellite (dotted) objects. The colours denote each volume-limited sample (see Table \ref{tab:z_sample}). The shaded areas denote the 1-$\sigma$ confidence levels on the model. For clarity, we only show it for the S0 and S4 samples.}
    \label{fig:central_satellites}
\end{figure}

We show the correlation functions and the best-fit models in Fig. \ref{fig:HOD_wprp}. For clarity, we shift the correlation functions by factors equal to $e^0$ for S0, $e^1$ for S1, $e^2$ for S2, $e^3$ for S3, $e^4$ for S4. The models in Eq. \ref{eq:HOD}, shown by the solid lines, represent the measurement well. The shaded areas show the 1- and 2-$\sigma$ uncertainties on the models. The total signal-to-noise ratio of the correlation function for each sample is 29.9, 39.4, 41.1, 35.4, and 29.7 from S0 to S4. The S4 sample (in violet) has the highest clustering signal, in agreement with the fact that it contains the largest fraction of the most massive objects. On the other hand, the S0 sample (in blue) probes lower mass clusters and groups and the clustering signal is lower, in agreement with expectations. \\
The best-fit parameters are reported with the respective priors in Table \ref{tab:parameter_HOD}. 
The $M_{\rm min}$ parameter decreases from 15.04$^{+0.61}_{-0.72}$ to 14.73$^{+0.93}_{-0.35}$ from the S4 to the S0 sample. It is in agreement with eROSITA probing the lower mass cluster population at low redshift. The $\sigma_{\log M}$ parameter decreases with redshift increasing. The S4 sample prefers a quick transition for the population of dark matter halos to host X-rays seen by eROSITA as a function of mass. For the S0 sample, the introduction of lower mass clusters and groups makes the transition smoother. The result is a larger $\sigma_{\log M}$, increasing from 0.57$^{+0.20}_{-0.37}$ to 0.82$^{+0.49}_{-0.46}$ from S4 to S0. The transition is nonetheless much sharper compared to AGN in the eROSITA Final Equatorial Depth Survey (eFEDS), where \citet[][]{Comparat2023A&Aagnclustering} measured $\sigma_{\log M}$=1.26. Similarly, the slope of the power law describing the satellites $\alpha_{\rm sat}$ is shallower for the low mass samples, varying from 1.25$^{+0.89}_{-0.54}$ to 1.67$^{+1.05}_{-1.08}$ from S0 to S4. For clusters and groups, the slope is steeper compared to eFEDS AGN \citep[$\alpha_{\rm sat}$=0.73,][]{Comparat2023A&Aagnclustering}, showing that according to our model the satellite population becomes quickly subsidiary as mass decreases in our study. The $M_{\rm sat}$ parameter assumes high values, that increase from 15.32$^{+0.34}_{-0.33}$ for S0 to 15.76$^{+0.42}_{-0.55}$ for S4. If satellite objects are rare, the HOD model will push the $M_{\rm sat}$ parameter to high values, fitting the correlation function mostly with signal from the central objects. We notice that most of the HOD parameters for different samples are compatible within uncertainties. Nonetheless, we see a trend for the median values from the faint low-$z$ population to the bright high-$z$ one in our volume-limited selection. Deeper data in future eRASS surveys with larger samples will reduce the uncertainty on the measure of the correlation functions due to the lower Poisson noise and help shed light on the satellite population of massive dark matter haloes in the clusters and groups regime. \\
We estimate the posterior distribution after accounting for the prior boundaries with the \texttt{GetDist}\footnote{\url{https://getdist.readthedocs.io/en/latest/}} software \citep[][]{Lewis2019arXiv_getdist}. The full marginalized posterior distributions are shown in Fig. \ref{fig:HOD_corner}. The five samples (S0, S1, S2, S3, S4) are shown in different colours (blue, orange, green, red, violet). The positive degeneracy between $M_{\rm min}$ and $\sigma_{\log M}$ is expected because a similar population of centrals may be described by a high cutoff mass but a broader transition or vice versa. The same holds for the negative degeneracy between $M_{\rm sat}$ and $\alpha_{\rm sat}$, with a steep power law anchored at large satellite masses explaining the same population of a shallower satellite function starting from lower masses. Together with $\sigma_{\log M}$, $\alpha_{\rm sat}$ is the HOD parameter with the broadest posterior.

We obtain the occupation distribution of central and satellite X-ray haloes seen by eROSITA as a function of mass for each volume-limited selected sample. The result is shown in Fig. \ref{fig:central_satellites}. The average number of centrals (satellites) in a halo of mass M is shown by the dotted (dashed) lines. According to our models, haloes more massive than 3.0 (6.5) $\times 10^{15}\,{\rm M}_\odot$ start hosting more than one satellite on average for the S0 (S4) sample. We obtain upper limits for the satellite fraction, which varies from <14.9$\%$ to <8.1$\%$ from S0 to S4. Our result confirms the general assumption that clusters detected in eRASS1 populate distinct haloes and the traditional halo mass function formalism for cluster abundance holds \citep{Ghirardini2024}.\\ 

\begin{figure}
    \centering
    \includegraphics[width=\columnwidth]{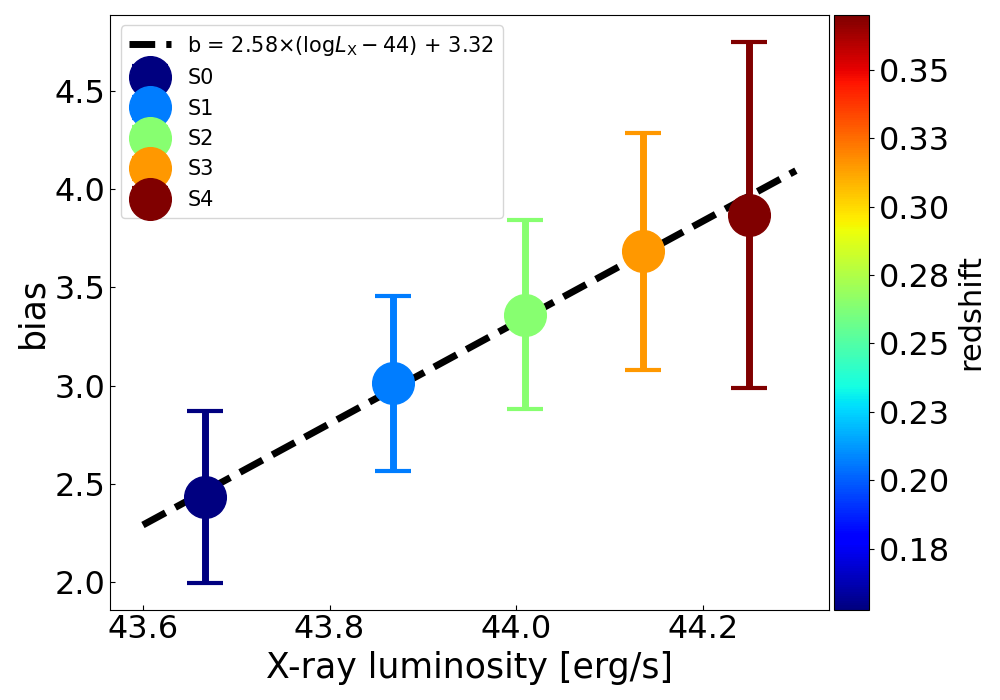}
    \caption{Large-scale halo bias as a function of average X-ray luminosity for the eRASS1 volume-limited samples. Each point is colour-coded by the average redshift of the sample. The black line shows the best-fit linear relation.}
    \label{fig:bias_Lx_z}
\end{figure}

\begin{figure}
    \centering
    \includegraphics[width=\columnwidth]{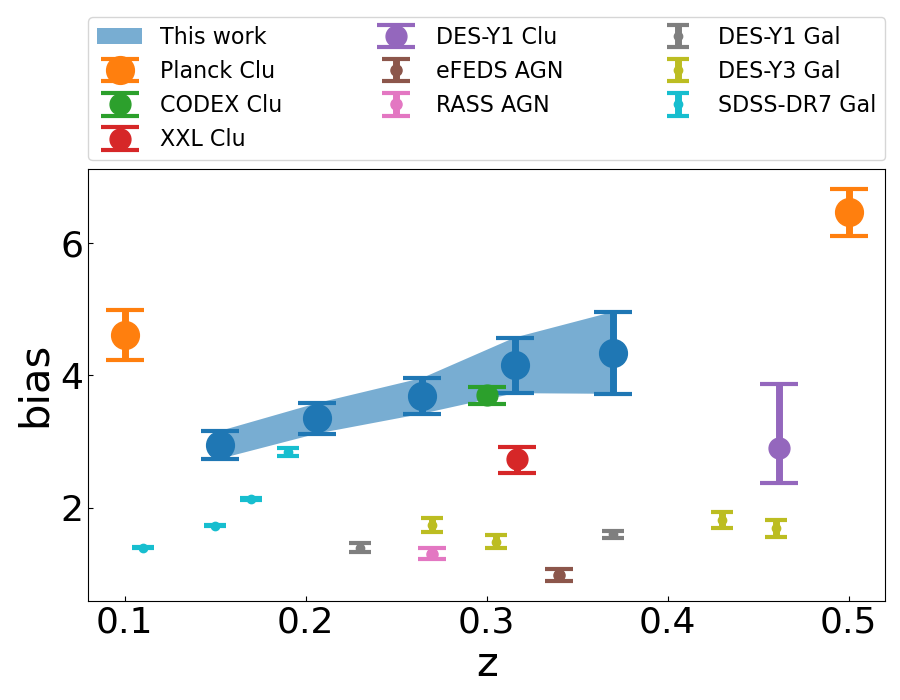}
    \caption{Comparison between the halo bias of eRASS1 clusters (in blue) with other experiments and probes from Planck clusters in orange \citep{Lesci2023A&A_planckclustering}, CODEX clusters in green \citep{Lindholm2021A&A...646A...8L}, XXL clusters in red \citep{Marulli2018A&A...620A...1M}, DES-Y1 clusters in purple \citep{Rykoff2016ApJS..224....1R}, eFEDS AGN in brown \citep{Comparat2023A&Aagnclustering}, RASS AGN in pink \citep{Krumpe2015ApJHOD}, DES-Y1 galaxies in grey \citep{Elvin-Poole2018PhRvD_DESY1clustering}, DES-Y3 galaxies in yellow \citep{Abbott2022PhRvD_DESY3}, and SDSS-DR7 galaxies selected in stellar mass in cyan \citep{ZuMandelbaum2015MNRAS_sdss_wprpggl}.
    }
    \label{fig:bias_other_probes}
\end{figure}

We find an increasing average halo bias as a function of mass and redshift. This is in agreement with a bottom-up structure formation scenario.
The bias changes by 47$\%$ from $b = 2.95 \pm 0.21$ in the S0 sample to $b = 4.34 \pm 0.62$ in S4. The bias as a function of the X-ray luminosity from \citet{Bulbul2024} for each sample is shown in Fig. \ref{fig:bias_Lx_z}. The panel is color-coded by the average redshift of each sample. We fit a linear relation between the bias and $\log L_X$ with \texttt{curve\_fit} in the form $b = m(\log L_{\rm X}-44) + q$. We find a slope of $m = 2.58\pm0.11$ and a zero point of $q = 3.32\pm 0.02$. The slope is much steeper in the cluster regime compared to AGN. In the clustering study of eFEDS, \citet[][]{Comparat2023A&Aagnclustering} combine their results with \citet[][]{Krumpe2015ApJHOD} and find a slope of 0.492. This is expected due to the strong dependency of the bias on halo mass. Our results are compatible with the bias models from \citet{Sheth2001MNRAS_ellipsoidal} respectively equal to 3.10 and 4.27, \citet[][]{Tinker2010ApJ_bias} equal to 3.14 and 4.56, \citet[][]{Comparat2017} equal to 3.12 and 4.46. The models from \citet[][]{Sheth1999MNRAS_LSbias} (2.81 and 3.88) and \citet{Bhattacharya2011} (2.68 and 3.83) underpredict our result. 
In Fig. \ref{fig:bias_other_probes} we show a comparison with bias values from other surveys and probes. For the DES-Y1 \texttt{redmapper} clusters \citep{Rykoff2016ApJS..224....1R} we derived an average halo bias with the model from \citet{Tinker2010ApJ_bias} using halo masses from the weak lensing richness-mass scaling relation from \citet{McClintock2019MNRAS_DESY1_redmapper_WL_SR}. Together with other cluster samples, our eRASS1 results show large bias values $b > 3$, compared to galaxies and AGN with values that are typically lower.  

\begin{figure}
    \centering
    \includegraphics[width=\columnwidth]{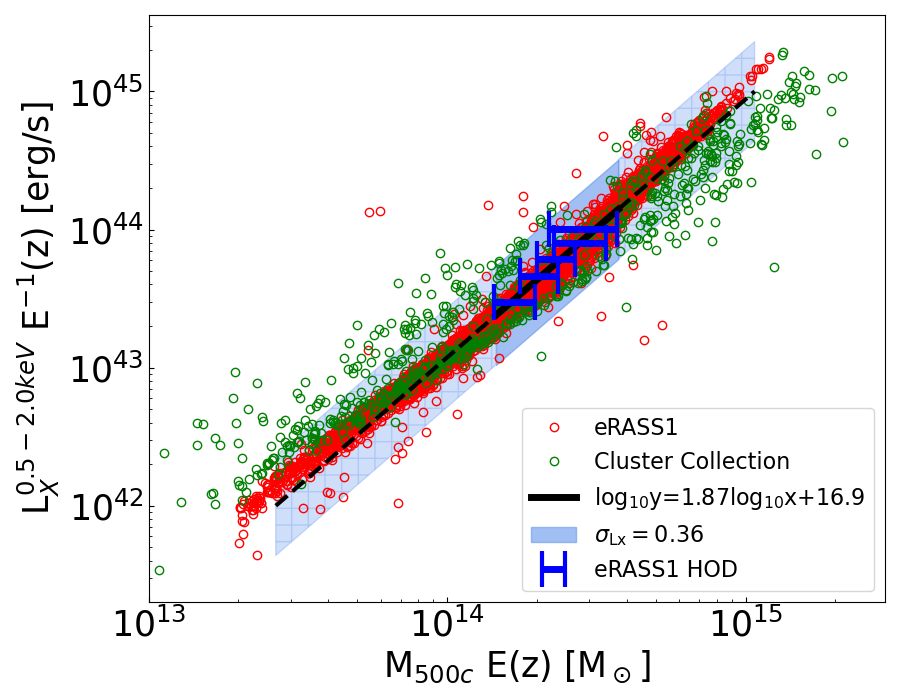}
    \caption{Scaling relation between halo mass and X-ray luminosity. We infer the halo mass from our best-fit HOD (in blue). The black line shows a linear fit to our data points. The green dots denote a collection of clusters from \citet[][]{Lovisari2015A&A...573A.118L_scalingrel}, \citet[][]{Mantz2016_scaling_relation}, \citet[][]{Schellenberger2017MNRAS.469.3738S_scalingrel}, \citet[][]{Adami2018A&A...620A...5A}, \citet[][]{Bulbul2019ApJ...871...50B_scalingrel}, \citet[][]{Lovisari2020ApJ...892..102L_scalingrel}, \citet[][]{2022A&A_LiuAng_eFEDS_clu}. The eRASS1 sample is denoted in red \citep{Bulbul2024}. The shaded area includes a scatter $\sigma_{\rm Lx} = 0.36$ around the mean relation, obtained by the halo abundance matching explained in Sect. \ref{subsec:HAM}. We mark the extrapolated scaling relation outside the parameter space sensitive to our HOD approach with the dashed black line and the hatched shaded area. 
    }
    \label{fig:Lx_M_SR}
\end{figure}

\begin{figure*}
    \centering
    \includegraphics[width=\columnwidth]{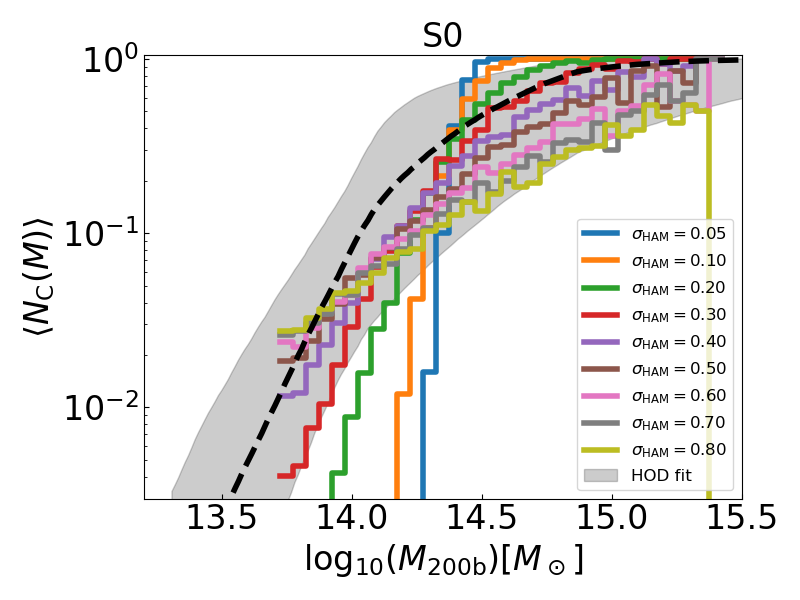}
    \includegraphics[width=\columnwidth]{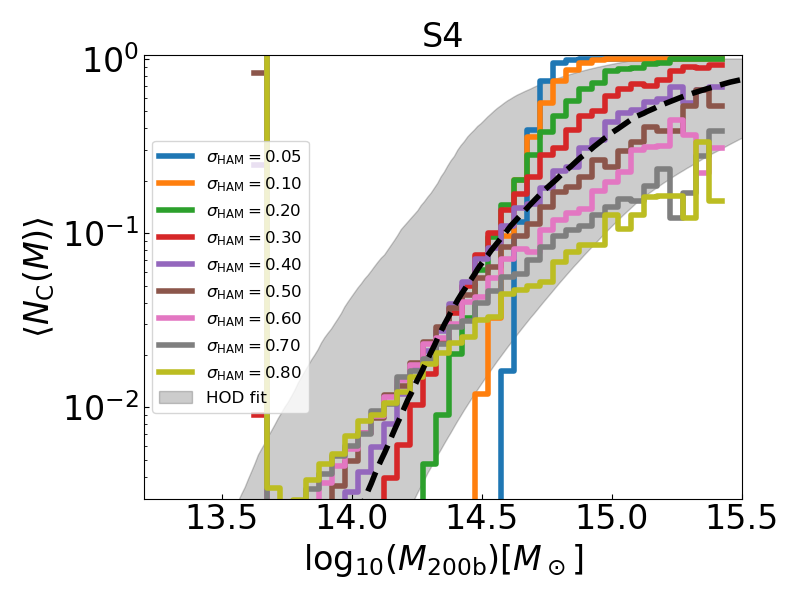}    \includegraphics[width=\columnwidth]{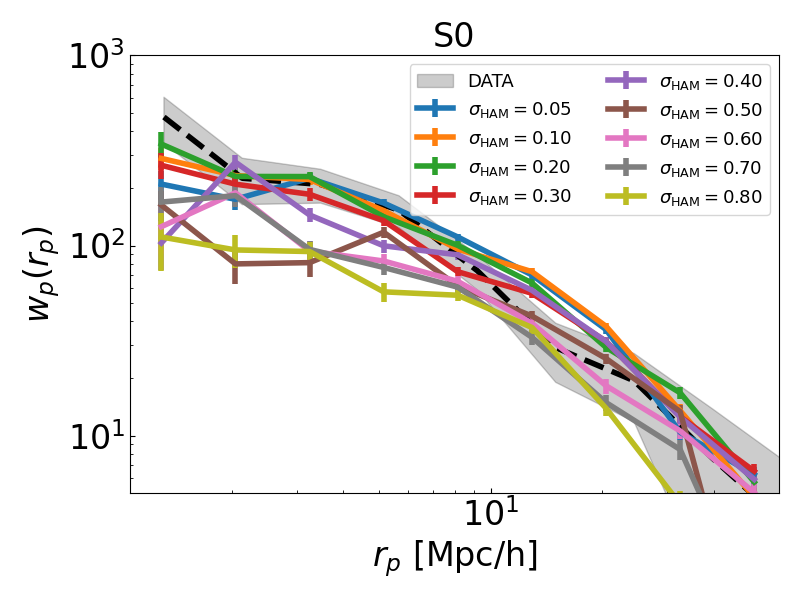}
    \includegraphics[width=\columnwidth]{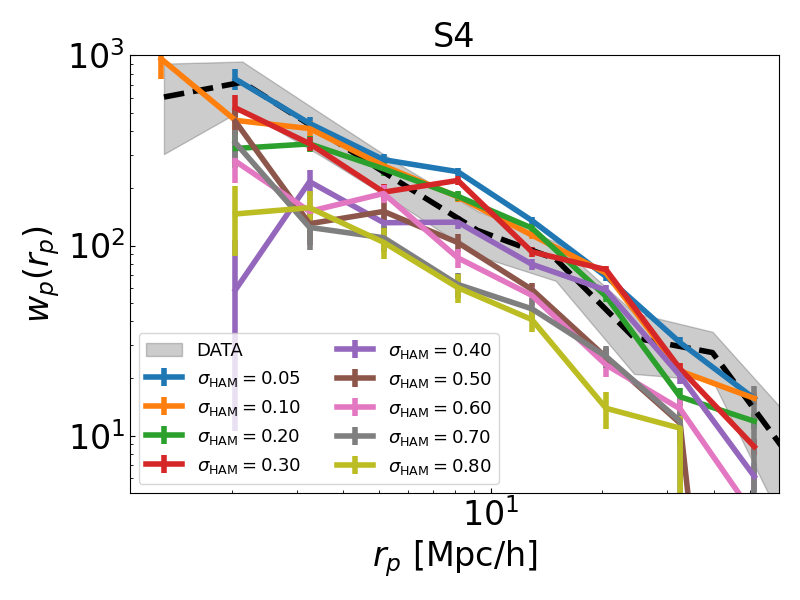}
    \caption{Comparison between the HOD results from Sect. \ref{subsec:HOD_result} and the abundance matching procedure. The top panels show the prediction for the number of central objects as a function of mass, the bottom ones show the projected correlation function. For clarity, we only show nine out of the hundred HAM predictions. The panels on the left refer to the S0 sample and the ones on the right to the S4 one. The shaded areas denote the results and measurements from the data, the coloured lines show the HAM predictions with different values of the scatter $\sigma_{\rm HAM}$.}
    \label{fig:HAM}
\end{figure*}

We infer the average halo mass for each volume-limited cluster sample from the best-fit HOD. We confirm that the low luminosity samples probe less massive haloes by inferring the halo mass populated by the five volume-limited samples from the HOD models. We find an average mass of $3.09 \pm 0.48 \times 10^{14}\,{\rm M_\odot}$ for S0, compared to $4.38 \pm 1.11 \times 10^{14}\,{\rm M_\odot}$ for S4. We convert $M_{\rm 200b}$ to $M_{\rm 500c}$ using the relations from \citet{Ragagnin2021MNRAShydromc}. In terms of $M_{\rm 500c}$ we obtain $1.58 \pm 0.25 \times 10^{14}\,{\rm M_\odot}$ for S0 and $2.42 \pm 0.61 \times 10^{14}\,{\rm M_\odot}$ for S4. The median value of the primary sample is $1.8 \times 10^{14}\,{\rm M_\odot}$ \citep{Bulbul2024}. Given the selection of our samples (see Table \ref{tab:z_sample}) we are well-aligned with the catalogue measurement. We then convert the average X-ray luminosity in the 0.2--2.3 keV band to 0.5-2.0 keV by computing conversion factors with \texttt{Xspec} \citep[][]{Arnaud1996ASPC..101...17A}, assuming an \textit{apec} model with the average temperature of the eRASS1 sample of kT=2.04 keV \citep{Bulbul2024}. The $L_X$--$M_{\rm 500c}$ scaling relation is shown in Fig. \ref{fig:Lx_M_SR}. Our data points are shown in blue. The green dots show a collection of clusters from \citet[][]{Lovisari2015A&A...573A.118L_scalingrel}, \citet[][]{Mantz2016_scaling_relation}, \citet[][]{Schellenberger2017MNRAS.469.3738S_scalingrel}, \citet[][]{Adami2018A&A...620A...5A}, \citet[][]{Bulbul2019ApJ...871...50B_scalingrel}, \citet[][]{Lovisari2020ApJ...892..102L_scalingrel}, and eFEDS \citep[][]{2022A&A_LiuAng_eFEDS_clu}. The red dots highlight the eRASS1 sample \citep{Bulbul2024}. The black line denotes a simple linear fit on our data and does not account for the selection function. We find a slope of 1.87$\pm$0.34 and a zero point of 16.9$\pm$4.2. Our result is compatible within 1 $\sigma$ with previous results in the literature \citep[]{Pratt2009A&A...498..361P, Lovisari2015A&A...573A.118L_scalingrel, Bulbul2019ApJ...871...50B_scalingrel, Capasso2020MNRAS_scalrel, Chiu2022A&A...661A..11C}. In addition, we combine constraints from $N_{\rm C}$ and $w_{\rm p}(r_{\rm p})$ to calibrate a HAM model using the eRASS1 clusters (see Sect. \ref{subsec:HAM}). We infer a scatter between X-ray luminosity and halo mass of $\sigma_{\rm Lx} = 0.36$, shown by the shaded area in Fig. \ref{fig:Lx_M_SR}. Our HOD approach anchors the X-ray-mass scaling relation with a new method, independent from mass calibration techniques using weak lensing of dynamical mass measurements, the results agree with previous studies. 

We compare the excess surface density prediction from our best-fit HOD models to lensing data in Appendix \ref{appendix:gglensing}. 

\subsection{Halo abundance matching predictions}
\label{subsec:HAM}

An alternative scheme to the HOD to model clustering measurements is halo abundance matching \citep[e.g.][]{Kravtsov2004ApJHOD, GuoZhengBehroozi_2016MNRAS.459.3040G}. 
It consists in using numerical N-body simulations (in a fixed cosmology) and devising a sub-halo selection procedure to match the observed number density of tracers as a function of redshift \citep{Rodriguez-TorresChuangPrada_2016MNRAS.460.1173R, Rodriguez-TorresComparatPrada_2017MNRAS.468..728R, FavoleComparatPrada_2016MNRAS.461.3421F, GuoYangRaichoor_2019ApJ...871..147G}. 

To construct an abundance matching model of X-ray selected clusters, we use the HugeMultiDarkPlanck (HMD) simulation \citep{Klypin2016}, a dark matter only simulation with a 4 $h^{-1}$ Gpc box size. We add a scatter $\sigma_{\rm HAM}$ around halo masses with values spanning from $\sigma_{\rm HAM} = 0$ to $1$ and intervals of 0.01, exploring the parameter space with 100 $\sigma_{\rm HAM}$ values. We then sort the values of mass and keep the most massive systems in the HAM samples to match the number density of sources in each eRASS1 volume selected sample (see Table \ref{tab:z_sample}). We obtain a HAM prediction for $N_{\rm C}$ by normalizing the mass histogram of each HAM sample with the mass histogram of the full HMD simulation. We also compute the projected correlation function for each HAM sample. Fig. \ref{fig:HAM} shows the result for S0 on the left and S4 on the right. The shaded areas denote the measurement of the eRASS1 correlation function and the HOD result, the HAM predictions with different values of the scatter $\sigma_{\rm HAM}$ are shown by the coloured lines. We find an overall good agreement between the HOD and HAM predictions. 

To find the optimal value of $\sigma_{\rm HAM}$, we compute two distances between the HAM prediction compared to best-fit HOD prediction for the number of centrals as a function of mass ($d_{\rm Nc}$) and the projected correlation function ($d_{\rm wp}$), according to:
\begin{align}
    d_{\rm Nc}(\sigma_{\rm HAM}) =& \frac{1}{N_{\rm M}} \sum_{\rm M} \frac{[N_{\rm HAM}(M,\sigma_{\rm HAM}) - N_{\rm HOD}^{\rm 50\%}(M)]^2}{[N_{\rm HOD}^{\rm 84\%}(M) - N_{\rm HOD}^{\rm 16\%}(M)]^2} \nonumber \\
    d_{\rm Wp}(\sigma_{\rm HAM}) =& \frac{1}{N_{\rm rp}} \sum_{\rm r_{\rm p}} \frac{[w_{\rm p,HAM}(r_{\rm p},\sigma_{\rm HAM}) - w_{\rm p,HOD}^{\rm 50\%}(r_{\rm p})]^2}{[w_{\rm p,HOD}^{\rm 84\%}(r_{\rm p}) - w_{\rm p,HOD}^{\rm 16\%}(r_{\rm p})]^2},
    \label{eq:dist_HAM}
\end{align}
where $N_{\rm M}$ ($N_{\rm rp}$) is the number of mass (projected separation) bins. We account for the uncertainty on the HOD model through the 16th and 84th percentiles of each distribution. Our formulation is very similar to the one adopted by \citet{Comparat2023A&Aagnclustering} for the HOD prediction, here we also include the correlation function term. We evaluate the allowed $\sigma_{\rm HAM}$ space as follows. For each $\sigma_{\rm HAM}$, we check whether both $d_{\rm Nc}(\sigma_{\rm HAM})$ and $d_{\rm Nc}(\sigma_{\rm HAM})$ are less than one, i.e. the difference between the best-fit HOD and the HAM prediction is within the HOD uncertainty. If this is the case, we mark the value of $\sigma_{\rm HAM}$ as allowed. We infer the optimal $\sigma_{\rm HAM}$ by averaging over the values allowed for each volume-limited sample. As Fig. \ref{fig:HAM} shows, the constraint from $w_{\rm p}(r_{\rm p})$ easily discards the highest values of $\sigma_{\rm HAM}$, whereas the one from $N_{\rm C}$ helps excluding the lowest ones. 

By construction, the HAM strictly depends on halo properties, in our case halo mass. Because of the X-ray luminosity cut in the eRASS1 volume-limited samples, our selection is not exactly mass-limited but includes the scatter of the observable (in this case $L_{\rm X}$) around halo mass. The $\sigma_{\rm HAM}$ parameter naturally accounts for it and is therefore related to the scatter in the scaling relation between $L_{\rm X}$ and halo mass \citep{GuoZhengBehroozi_2016MNRAS.459.3040G}. In other words, a HAM with $\sigma_{\rm HAM} = 0$ would reflect an observable capable of perfectly tracing halo mass. It is then possible to deduce the average X-ray luminosity scatter at fixed mass by accounting for the slope $m = 1.87$ of the $L_{\rm X} - M_{\rm 500c}$ scaling relation from Sect. \ref{subsec:HOD_result} and assuming that such scatter is lognormal. The end result is $\sigma_{\rm Lx} = \dfrac{m\sigma_{\rm HAM}}{\sqrt{2}}$ \citep[see][]{Zheng2007ApJ_HOD, GuoZhengBehroozi_2016MNRAS.459.3040G}. We collect the results in Table \ref{tab:sigma_ham}.

\begin{table}[]
    \caption{Optimal values of the scatter parameter $\sigma_{\rm HAM}$ for each volume-limited sample.}
    \label{tab:sigma_ham}
    \centering
    \begin{tabular}{| c | c | c |}
    \hline
    \hline
    \rule{0pt}{2.2ex}
    Sample & $\sigma_{\rm HAM}$ & $\sigma_{\rm Lx}$ \\
    \hline
    \rule{0pt}{2.2ex}
    S0 & $0.44 \pm 0.23$ & $0.59 \pm 0.34$ \\
    \hline
    \rule{0pt}{2.2ex}
    S1 & $0.23 \pm 0.13$ & $0.31 \pm 0.18$ \\
    \hline
    \rule{0pt}{2.2ex}
    S2 & $0.26 \pm 0.15$ & $ 0.34 \pm 0.20$ \\
    \hline
    \rule{0pt}{2.2ex}
    S3 & $0.27 \pm 0.16$ & $0.35 \pm 0.21$ \\
    \hline
    \rule{0pt}{2.2ex}
    S4 & $0.17 \pm 0.09$ & $0.23 \pm 0.13$ \\
    \hline    
    \end{tabular}
    \tablefoot{We additionally infer the scatter between X-ray luminosity and halo mass $\sigma_{\rm Lx}$ between X-ray luminosity and halo mass from the HAM procedure as explained in the text.
    }
\end{table}

Albeit our constraints are broad and the $\sigma_{\rm HAM}$ are compatible within error bars with each other, we find that on average it decreases by about 60$\%$ from $0.44$ for S0 to $0.17$ for S4. This is in agreement with our HOD results and the sharper selection for luminous massive clusters. 
By averaging over the individual inferred luminosity-mass scatter on each sample, we find $\langle \sigma_{\rm Lx} \rangle = 0.36 \pm 0.12$ for eRASS1. Therefore, we add such scatter to the scaling relation obtained in Sect. \ref{subsec:HOD_result} (Fig. \ref{fig:Lx_M_SR}). Part of this is scatter resides in the intrinsic nature of clusters, for example, different core properties are reflected in varying luminosities at fixed halo mass. In addition, the luminosity selection from the eROSITA survey adds an additional component to the scatter. This description is consistent with our result in comparison to the X-ray cluster model from \citet{Comparat2020Xray_simulation}, who find an intrinsic scatter of 0.21. The model does not include the secondary scatter component due to the survey selection function. Finally, our result encompasses the cluster samples from different surveys with good agreement (see Fig. \ref{fig:Lx_M_SR}). In particular, \citet{Bulbul2019ApJ...871...50B_scalingrel} found a scatter of $0.27^{\rm + 0.08}_{\rm - 0.12}$ for an SZ-selected sample around the mean scaling relation, compatible with our result within 1$\sigma$.

In summary, the HAM procedure starts from a set of dark matter haloes generated by a Universe with \citet{Planck2016A&A..cosmopars} cosmology, the one assumed by the HMD simulation. The fact that the $\sigma_{\rm HAM}$ obtained with our method is in agreement with the literature means that this study with eRASS1 clusters does not exclude such set of cosmological parameters.
From the HAM perspective, we conclude that our results are compatible with the \citet{Planck2016A&A..cosmopars} cosmology.

\subsection{Cosmological implications}
\label{subsec:cosmo_results}

\begin{figure}
    \centering
    \includegraphics[width=\columnwidth]{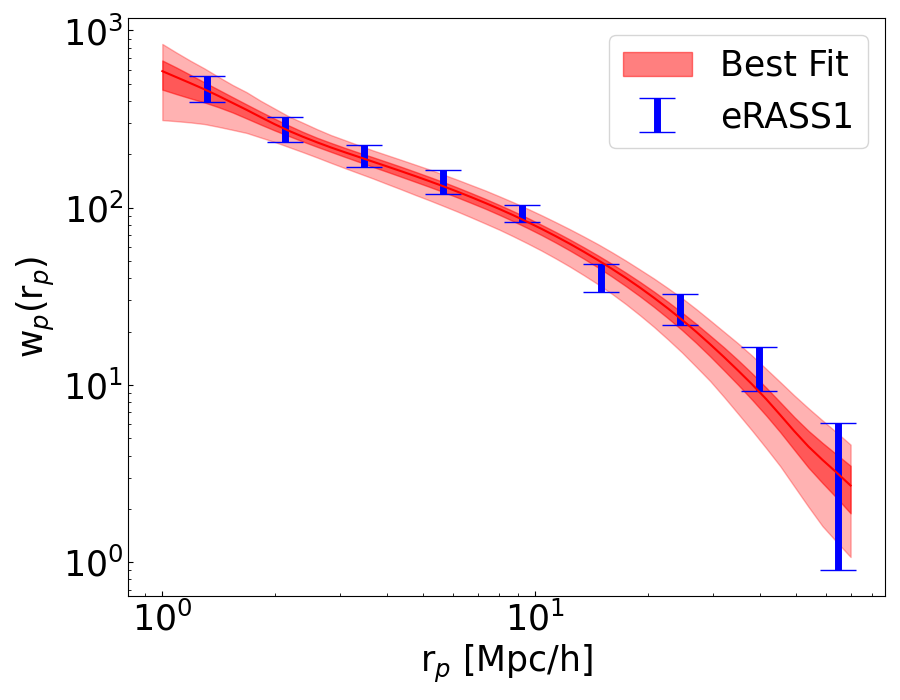}
    \caption{Projected correlation functions of the eRASS1 clusters. The error bars show the measurement, the solid line is the best fit cosmological model, and the shaded areas denote the 1-$\sigma$ and 2-$\sigma$ uncertainty on the model.}
    \label{fig:wprp_cosmo}
\end{figure}

For the inference of cosmological parameters, we focus on a flux-limited sample of 6493 sources with the same luminosity cut of S0. We still consider the redshift range 0.1<z<0.6. The efficient removal of contaminants (see Sect. \ref{sec:data}), makes it suitable also for cosmological studies. The methodology is explained in Sect. \ref{subsubsec:Cosmo}. Thanks to the larger number of clusters in this sample compared to the volume-limited ones, the signal-to-noise ratio is larger and reaches 50.1. Figure \ref{fig:wprp_cosmo} shows the measured correlation function with error bars in blue and the best-fit cosmological model in red. The shaded areas denote the 1- and 2-$\sigma$ uncertainty on the model.

\begin{table}[]
    \caption{Priors and posteriors on the cosmology and large-scale halo bias parameters.}
    \label{tab:parameter_cosmo_bias}
    \centering
    \begin{tabular}{| c | c | c |}
    \hline
    \hline
    Parameter & Prior & Posterior\\
    \hline
    \rule{0pt}{2.2ex}
        $\omega_{\rm c}$ & $\mathcal{U}(0.001, 0.3)$ &  0.12$^{+0.03}_{-0.02}$ \\
        \rule{0pt}{2.2ex}
        $\ln 10^{10}A_{\rm s}$ & $\mathcal{U}(1.0, 5.0)$ & 2.92$^{+0.55}_{-0.53}$ \\
        \rule{0pt}{2.2ex}
        $\omega_{\rm DE}$ & $\mathcal{U}(0.2, 0.4)$ & 0.29$_{-0.07}^{+0.07}$ \\
        \rule{0pt}{2.2ex}
        $M_{\rm min}$ & $\mathcal{U}(14.0, 15.5)$ & 14.75$^{+0.25}_{-0.26}$ \\
    \hline
        \rule{0pt}{2.2ex}
        $b$ & - & 3.63$^{+1.02}_{-0.85}$ \\  \rule{0pt}{2.2ex}
        $\sigma_{\rm 12}$ & - & 0.77$^{+0.22}_{-0.16}$ \\
        \rule{0pt}{2.2ex}
        $\Omega_{\rm M}$ & - & 0.32$^{+0.04}_{-0.03}$\\
        \rule{0pt}{2.2ex}
        $\sigma_{\rm 8}$ & - & 0.76$^{+0.22}_{-0.16}$ \\
        $H_0$ & - & 65.5$^{+6.8}_{-5.6}$ \\  
    \hline    
    \end{tabular}
    \tablefoot{The upper part of the table reports the parameters directly sampled by our fitting method, while the bottom part shows inferred parameters. $\mathcal{U}({\rm min}, {\rm max})$ indicates a uniform prior between min, max values. 
    }
\end{table}

\begin{figure}
    \centering
    \includegraphics[width=\columnwidth]{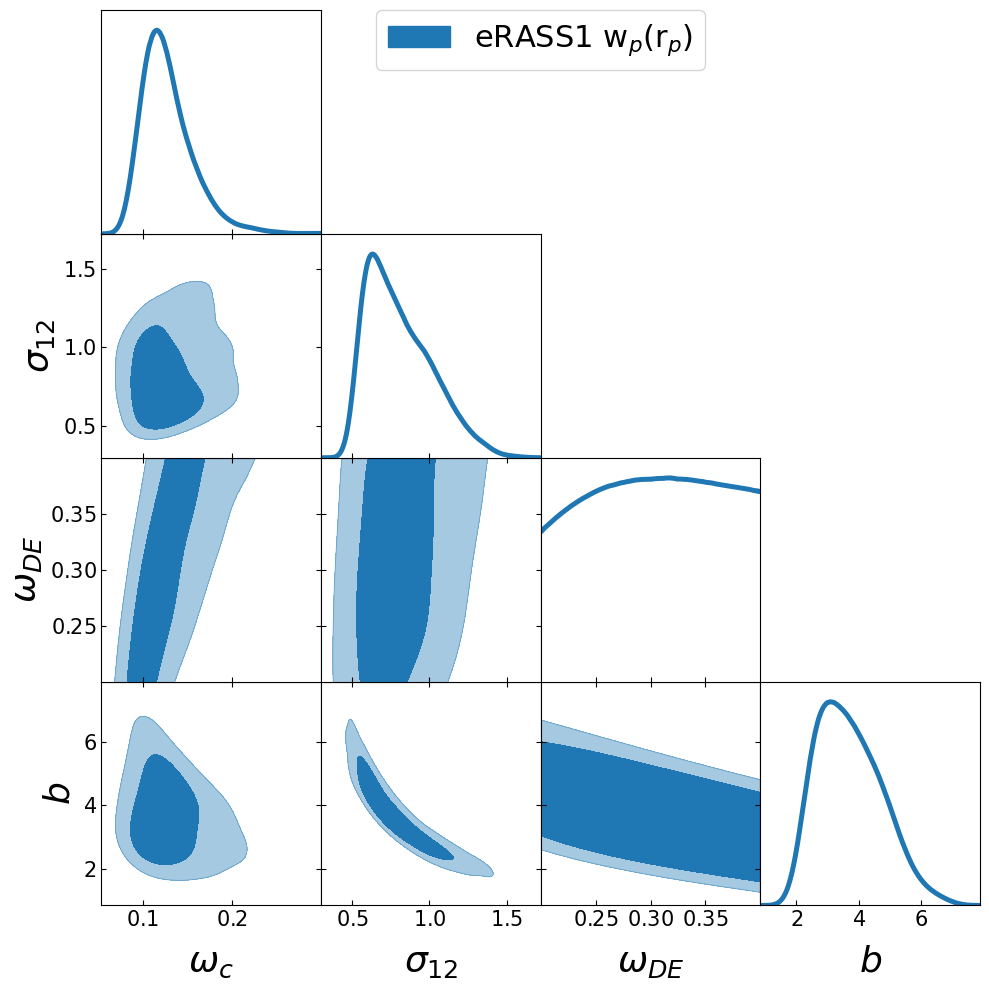}
    \caption{Marginalized posterior distributions of the best fit cosmological parameters. The 1-$\sigma$ and 2-$\sigma$ confidence levels of the posteriors are shown as filled 2D contours. The model is explained in Sect. \ref{subsubsec:Cosmo}. The parameters are reported in Table \ref{tab:parameter_cosmo_bias}.}
    \label{fig:posterior_HOD_cosmo}
\end{figure}

\begin{figure*}
    \centering
    \includegraphics[width=\columnwidth]{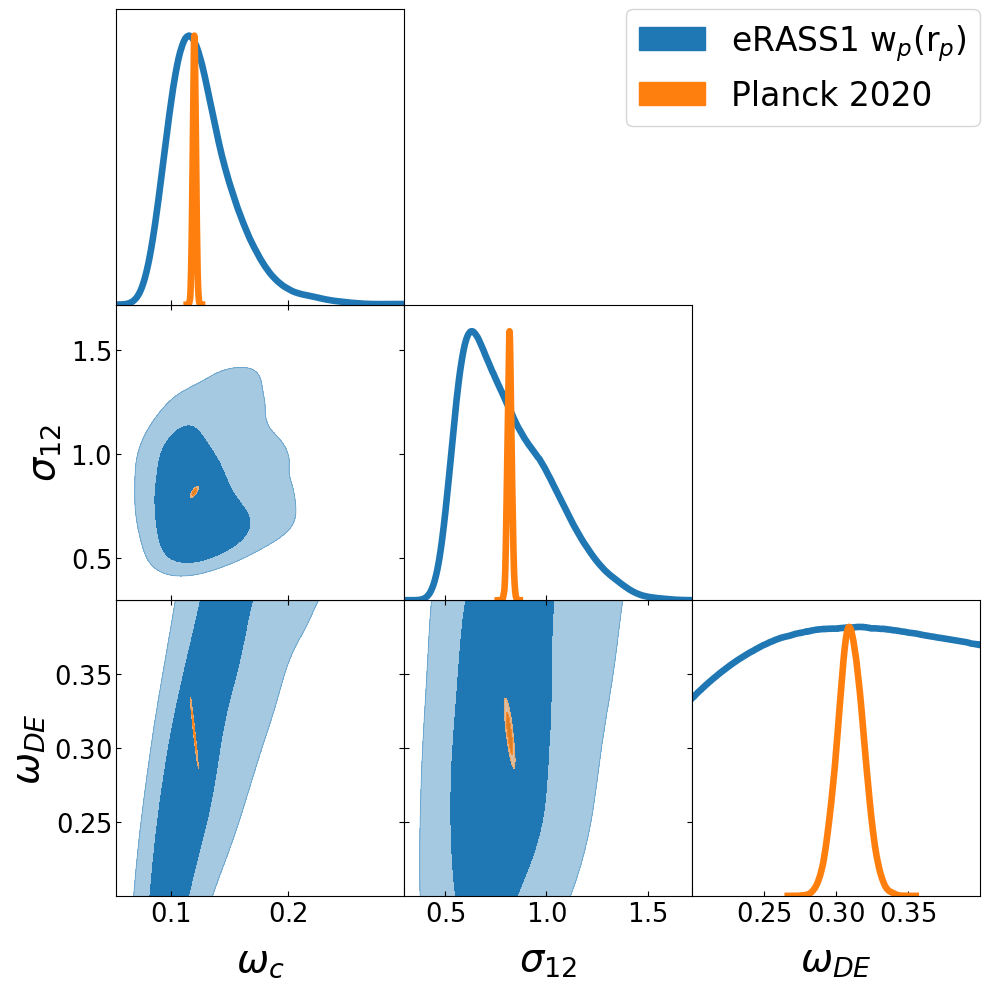}
    \includegraphics[width=\columnwidth]{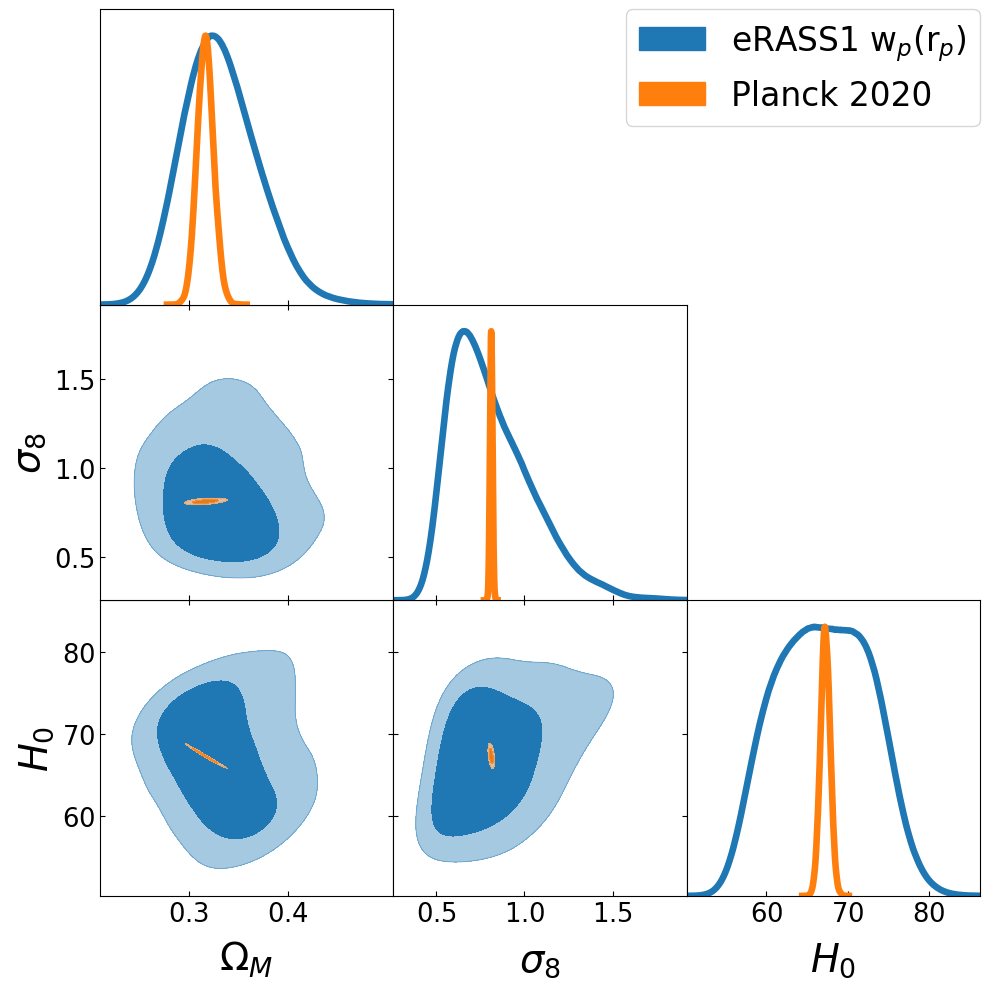}
    \caption{Marginalized posterior distributions of the best fit cosmological parameters. The 1-$\sigma$ and 2-$\sigma$ confidence levels of the posteriors are shown as filled 2D contours. The model is explained in Sect. \ref{subsubsec:Cosmo}. The parameters are reported in Table \ref{tab:parameter_cosmo_bias}. The left-hand panel shows $h$-independent parameters: the physical cold dark matter density $\omega_{\rm c}$, the clustering amplitude $\sigma_{\rm 12}$, and the physical dark energy density $\omega_{\rm DE}$. The right-hand panel shows the traditional inferred parameters: the matter density $\Omega_{\rm M}$, the amplitude $\sigma_{\rm 8}$ and the Hubble constant $H_0$. The constraints from the flux-limited eRASS1 sample are shown in blue, the CMB results from \citet[][]{Planck2020A&A...641A...6P} in orange.}
    \label{fig:posterior_cosmo_eRASS1}
\end{figure*}

We interpret our results in terms of the clustering amplitude $\sigma_{\rm 12}$ \citep[][]{Sanchez2022MNRAS_evomapping}, encoding the variance of the density field smoothed on scales of 12 Mpc. It is defined according to 
\begin{equation}
    \sigma_{\rm 12}^2 = \frac{1}{2\pi^2}\int dk k^2 P_{\rm lin}(z=0,k)\Tilde{W}(k\times12\text{Mpc}),
    \label{eq:sigma12}
\end{equation}
where $P_{lin}$ is the linear matter power spectrum, and $\Tilde{W}$ is the Fourier transform of a top-hat filter with a radius of 12 Mpc. $\sigma_{\rm 12}$ has the advantage of depending on the Hubble parameter only through the amplitude of the power spectrum, not on the reference smoothing scale. This allows us to disentangle the dependency of the clustering amplitude on $A_{\rm s}$ through $\sigma_{\rm 12}$ from the one on the reference scale encoded in $h$. \\
The prior and the posteriors for the cosmological parameters are reported in Table \ref{tab:parameter_cosmo_bias}. We obtain the large-scale halo bias as a derived parameter from the HOD formalism. We measure $\omega_{\rm c}=0.12^{+0.03}_{-0.02}$, 
$\ln 10^{10}A_{\rm s}=2.92^{+0.55}_{-0.53}$, and $M_{\rm min}=14.74^{+0.28}_{-0.24}$. We infer the derived parameters $b=3.63^{+1.02}_{-0.85}$, $\sigma_{\rm 12}=0.77^{+0.22}_{-0.16}$, $\Omega_{\rm M}=0.32_{-0.03}^{+0.04}$, and $\sigma_{\rm 8}=0.76^{+0.22}_{-0.16}$. Our constraining power on $\omega_{\rm DE}$ is weak, and the best-fit value we get from the posterior fills the prior. Therefore, the $H_0$ constraint is conservative. Our results are compatible within 1-$\sigma$ with state-of-the-art results from the cosmic microwave background (CMB) analysis \citep[][]{Planck2020A&A...641A...6P} and provide an independent test for the vanilla $\Lambda$CDM cosmological model. \\
The panels in Fig. \ref{fig:posterior_HOD_cosmo} show the full marginalised posterior distribution. As expected, we find a strong negative degeneracy between the large-scale-halo bias and the clustering amplitude $\sigma_{\rm 12}$. The relative correlation coefficient is -0.91. Indeed, they are both directly related to the overall normalization of the correlation function. Because of this aspect, the constraining power on $\sigma_{\rm 12}$ (and similarly for $\sigma_{\rm 8}$) is weak, which also makes the $\omega_{\rm c}$--$\sigma_{\rm 12}$ 2D posterior distribution broad. We compare the $h$-independent cosmological parameters and the traditional ones in Fig. \ref{fig:posterior_cosmo_eRASS1}. The eRASS1 clustering constraints are shown in blue, and the CMB ones from \citet[][]{Planck2020A&A...641A...6P} in orange. We find a positive degeneracy between $\omega_{\rm c}$ and $\omega_{\rm DE}$, with a correlation coefficient of 0.61. The same behaviour was obtained by \citet[][]{Semenaite2022MNRAS_eBOSS} in a clustering study of the Extended Baryon Oscillation Spectroscopic Survey (eBOSS). Indeed, the same LSS at redshift about 0.3 can be generated in a Universe with a high physical matter density facilitating halo collapse and a high amount of dark energy contrasting the collapse, or vice versa. The difference between such scenarios resides in the expansion factor regulated by $H_0$. The correlation between $\Omega_{\rm M}$ and $H_0$ is in fact negative, with a coefficient of -0.27. 

\begin{figure}
    \centering
    \includegraphics[width=\columnwidth]{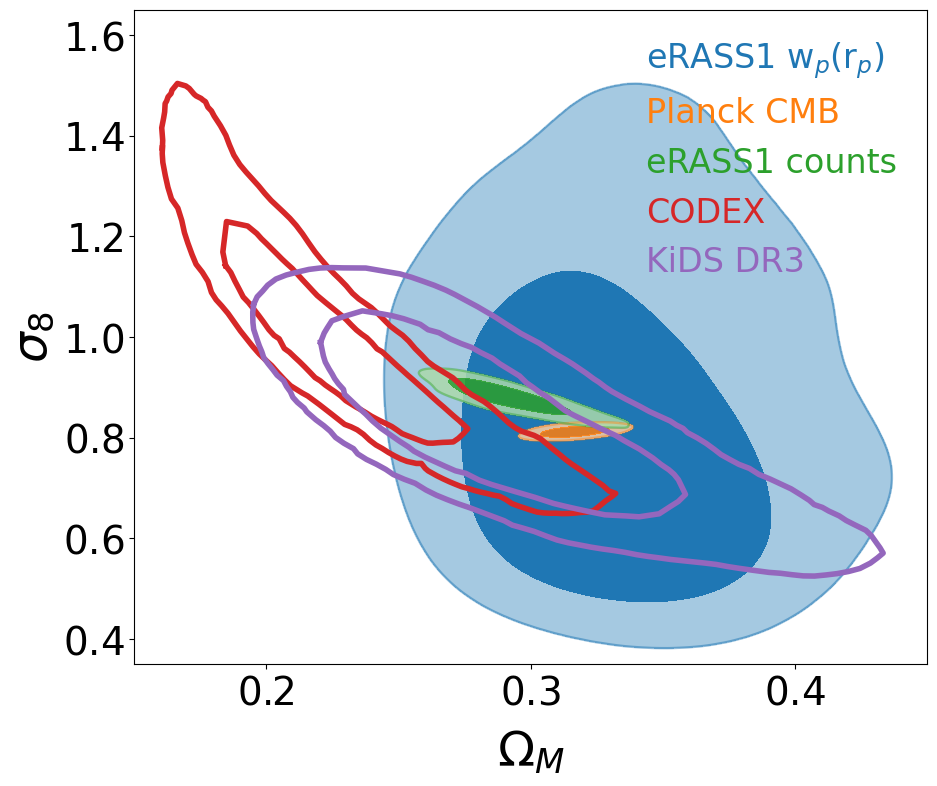}
    \caption{Derived constraints on $\Omega_{\rm M}$ and $\sigma_{\rm 8}$. Our result with the eRASS1 cluster clustering is shown in blue, the result from \citet[][]{Planck2020A&A...641A...6P} (EEE TTT, lowL, lowE) in orange, eRASS1 cluster abundance \citep{Ghirardini2024} in green, CODEX \citep[][]{Lindholm2021A&A...646A...8L} in red, and KiDS-DR3 \citep[][]{Lesci2022A&A_clustering} in violet.}
    \label{fig:cosmo_contours}
\end{figure}

We compare our results to other cosmological experiments involving the clustering of galaxy clusters in Fig. \ref{fig:cosmo_contours}. It shows our constraints on $\Omega_{\rm M}$ and $\sigma_{\rm 8}$ in blue, the result from CODEX clusters \citep[][]{Lindholm2021A&A...646A...8L} in red, and from AMICO KiDS-DR3 clusters \citep[][]{Lesci2022A&A_clustering} in violet. The CMB constraint from \citep[][]{Planck2020A&A...641A...6P} is shown in orange. The green area denotes the constraints from eRASS1 cluster abundance \citep{Ghirardini2024}. Our results are compatible well within 1$\sigma$. Although our model does not rely on any large-scale halo bias model, our result is comparable to previous clustering studies with clusters of galaxies. The larger amount of clusters available for our eRASS1 experiment (6493) allows us to obtain tighter constraints on $\Omega_{\rm M}$, with about 10$\%$ uncertainty compared to CODEX (1892) and KiDS-DR3 (4934), that had 16$\%$ uncertainty. Conversely, our free-bias HOD approach, with no assumptions on cluster masses and scaling relations makes the constraints less prone to systematics, but looser on $\sigma_{\rm 8}$, also due to the weak constraining power on $\omega_{\rm DE}$ and its degeneracy with both matter density and clustering amplitude. \\  
The individual constraints from eRASS1 cluster clustering are not powerful enough to probe cosmological tensions such as the S8 or $H_0$ ones \citep[][]{Verde2019NatAs_tensions}. The deeper data from future eRASS surveys will increase the number of clusters and reduce the uncertainty of the correlation function. Thanks to the combination with cluster counts \citep{Ghirardini2024}, eROSITA will measure cosmological parameters with per cent level accuracy \citep[][]{Merloni2012, Pillepich2018}.

\section{Conclusions}
\label{sec:conclusions}

We summarize our results in Sect. \ref{subsec:summary} and provide a future outlook in Sect. \ref{subsec:discussion}.

\subsection{Summary}
\label{subsec:summary}

The ability of eROSITA to effectively detect extended extra-galactic sources yields large samples of galaxy clusters. The eROSITA telescope has observed 12\,247 cluster candidates during its first all-sky survey \citep{Bulbul2024, Kluge2024}. Such a large sample is suitable for exploratory studies of their clustering, allowing us to learn about their distribution within the large-scale structure of the Universe and the dark matter halo population hosting such objects. In addition, clustering measurements provide individual cosmological constraints, independently from traditional cluster abundance experiments.

We studied the clustering of galaxy clusters detected in the first eROSITA all-sky survey (eRASS1). We generated a catalogue of random points which traces the same density and redshift distributions of the real data by construction, thanks to filtering according to the eRASS1 selection function (see Fig. \ref{fig:sky_plot} and \ref{fig:dndz}). We applied a volume-limited selection in X-ray luminosity and redshift, by discarding clusters with $L_X$ lower than the one inferred from a reference flux $F_{\rm X,lim}=4\times 10^{-14}\,{\rm erg/s/cm}^2$ at the upper redshift limit of each sample. We obtained five different samples, whose properties are reported in Table \ref{tab:z_sample}. We measured the two-point autocorrelation function of each sample (see Eq. \ref{eq:LS93}). We computed covariance matrices by a jackknife method on the eRASS1 data and on a set of 1000 independent GLASS simulations (see Eq. \ref{eq:cov_mat} and Fig. \ref{fig:correlation_matrix}). 

We interpreted our measurement with a HOD approach (see Fig. \ref{fig:HOD_wprp}). To our knowledge, this is the first attempt to constrain HOD parameters for X-ray selected clusters. We model the eRASS1 clusters incompleteness with the eRASS1 digital twin from \citep[][]{Seppi2022A&A_eRASS1sim}. We found that bright clusters probe high-mass haloes, as expected. The $M_{\rm min}$ parameter shifts from 14.73$^{+0.93}_{-0.35}$ for S0 to 15.04$^{+0.61}_{-0.72}$ for S4. The large-scale halo bias consequently decreases from 4.34$\pm$0.62 for S4 to 2.95$\pm$0.21 for S0. We found that satellites populate a larger fraction of haloes in the low-luminosity samples. The $M_{\rm sat}$ parameter changes from 15.32$^{+0.34}_{-0.33}$ to 15.76$^{+0.42}_{-0.55}$ (see also Fig. \ref{fig:central_satellites}).  The high-mass samples prefer a sharp transition for central objects populating haloes, with $\sigma_{\log M} =0.57_{-0.34}^{+0.29}$ for S4. When probing lower-mass sources, eROSITA detects more satellite objects, and the transition is milder, with $\sigma_{\log M}=0.82_{-0.49}^{+0.43}$ for S0 (see Fig. \ref{fig:HOD_corner} for the full posterior distribution). As a consequence, the fraction of satellites (Eq. \ref{eq:frac_sat}) reaches higher upper limits for low L$_{\rm X}$ samples (from <8.1$\%$ to <14.9$\%$). The priors, posteriors, and derived HOD parameters are reported in Table \ref{tab:parameter_HOD}. Using the best-fit HOD, we studied the relation between $L_X$ and the bias in Fig. \ref{fig:bias_Lx_z}, and the scaling relation between $L_X$ and $M_{\rm 500c}$ in Fig. \ref{fig:Lx_M_SR}. The results agree with previous studies. We used a HAM approach to infer the scatter of X-ray luminosity at fixed halo mass, combining the five volume-limited samples we obtain $\langle \sigma_{\rm Lx} \rangle = 0.36$, in agreement with the literature. Our model agrees with measurements of the excess surface density obtained from the cross-correlation between the cluster positions and the galaxy shapes from KiDS data (Fig. \ref{fig:deltasigma}).

We focus on a larger flux-limited sample to study the cosmological implications of our clustering measurement. It consists of 6493 sources. We model the large-scale halo bias within the HOD framework without assuming a fiducial model. We follow the evolution mapping approach from \citet[][]{Sanchez2022MNRAS_evomapping} and interpret our results in an $h$-independent way. We fix all the cosmological parameters affecting the shape of the power spectrum, except for the physical cold dark matter density $\omega_{\rm c}$. We leave the evolution parameters $A_{\rm s}$ and $\omega_{\rm DE}$ free, together with $M_{\rm min}$. We use the other HOD variables as nuisance parameters. We obtain constraints on $\omega_{\rm c}=0.12^{+0.03}_{-0.02}$ and $\sigma_{\rm 12}=0.77^{+0.22}_{-0.16}$. The average bias of the flux-limited sample selected for cosmology is $b=3.63^{+1.02}_{-0.85}$. We derive constraints on $\Omega_{\rm M}=0.32_{-0.03}^{+0.04}$, $\sigma_{\rm 8}=0.76^{+0.22}_{-0.16}$. Our results are in excellent agreement with eRASS1 cluster abundance \citep{Ghirardini2024}. The full posterior distribution is shown in Fig. \ref{fig:posterior_cosmo_eRASS1}. Our results are compatible with similar clustering studies of galaxy clusters \citep[][]{Lindholm2021A&A...646A...8L, Lesci2022A&A_clustering} and CMB \citep[][]{Planck2020A&A...641A...6P}.

\subsection{Outlook}
\label{subsec:discussion}

The future eROSITA all-sky surveys, thanks to the deeper data, will provide a more complete view of satellite groups to massive dark matter haloes, allowing us to probe clustering measurements on smaller scales and obtain better constraints on the satellite population from HOD modelling. Thanks to the development of new high-resolution mock Universes, a detailed comparison with simulations will then be possible.
The next generation of cosmological simulations with, such as FLAMINGO \citep[][]{Schaye2023MNRAS_flamingo}, MilleniumTNG \citep[][]{Pakmor2023MNRAS_MillenniumTNG}, and Cluster-TNG \citep[][]{Nelson2023arXiv231106338N_TNG-Cluster} will help shed light on the fraction of satellites in the whole cluster population from the theoretical point of view.

A better understanding of the population of satellite haloes will also benefit from precise measurement and modelling of the subhalo mass function. From the perspective of simulations, the definition of a subhalo is not trivial. \citet[][]{Diemer2023_hauntedhalos} showed that traditional halo finders may fail in the subhalo identification due to tidal distortions and low particle resolution. They track subhalo particles in the simulations and obtain a better view of the whole subhalo population, which impacts the clustering measurements by more than 20$\%$ on small scales below 1 Mpc also for haloes more massive than 10$^{13}$ M$_\odot$. Such an approach will ultimately offer a more accurate prediction for the expected HOD parameters such as $M_{\rm sat}$, $\alpha_{\rm sat}$, and the satellite fraction.

Another key ingredient for an accurate prediction of small-scale clustering is a reliable modelling of the additional suppression of the non-linear power spectrum due to baryons. Different works in the past few years provide ad-hoc prescriptions from the comparison of dark matter-only simulations with parent runs including baryons, also as a function of the baryonic physics implementation \citep[see e.g.,][]{Arico2020MNRAS_baryonsPk, Salcido2023MNRAS_baryonPk, Grandis2023arXiv_Pk}.

Multiple works showed that the modelling of the excess surface density (ESD) improves the HOD modelling \citep[see, e.g.][for AGN and galaxies]{Comparat2023A&Aagnclustering, More2023}. Accurate models for clusters require accounting for mis-centering effects \citep[][]{Zhang2019MNRAS.487.2578Z}, cluster member contamination \citep[][]{Varga10.1093_clumemcont}, and redshift uncertainties \citep[see discussion by][]{Rykoff2014redmapper}. Future developments on this side will allow us to use ESD measurements not only as a consistency check (see Sect. \ref{appendix:gglensing}) but also as an additional term in the HOD likelihood together with clustering and improve the HOD constraints, making the best out of combined statistics for cosmological cluster studies \citep[][]{To2021_DES_cosmology}.

Accurate and precise redshift measurements are key to studying the BAO feature using galaxies and clusters as probes of the Large-Scale Structure \citep[][]{Comparat2013MNRAS_BAO_sel}. The current quality provided by photometric redshifts is not sufficient \citep{Kluge2024}. In the future, spectroscopic follow-up of eROSITA clusters from SDSS-V \citep[][]{Kollmeier2017_SDSS5} and 4MOST \citep[][]{deJong2011_4MOST, Finoguenov2019Msngr} will allow us to go beyond the measurement of projected clustering, enabling a detailed study of the 3D cluster clustering, redshift space distortions, and BAO modelling.

Clustering studies of the deeper eROSITA surveys such as eRASS:4 will allow us to fit cosmological parameters for different samples as a function of redshift, with better handling of the strong degeneracy between halo bias and clustering amplitude. The simultaneous fitting of the clustering signal and the cluster abundance \citep[][]{Ghirardini2024} will allow an additional HOD constraint on the cluster number density, anchoring the clustering amplitude and partially solving the degeneracy with the halo bias. Similarly, the sensitivity to $\omega_{\rm DE}$ will improve, together with the constraints on $H_{\rm 0}$. This will allow us to explore a larger portion of the cosmological parameter space and highlight the different degeneracy for the halo bias with $\sigma_{\rm 8}$ and $\sigma_{\rm 12}$. In addition, we will potentially go beyond the traditional $\Lambda$CDM model and provide constraints on primordial non-gaussianity \citep[][]{Sartoris2010MNRAS_nongauss}.

Finally, future studies with larger catalogues and higher number densities will require a precise estimate of the covariance matrix. The generation of end-to-end mocks is computationally very expensive. Such studies will benefit from fast and accurate estimates of the covariance matrix \citep[][]{Trusov2023_covmat}, especially if the clustering is combined with external probes, such as cluster counts, where both experiments share common information \citep[][]{Fumagalli2023arXiv_cosmosdss}.

\section*{Acknowledgements}

R. Seppi thanks S. More for advice on the \texttt{AUM} software.
NC is supported by CNES.

This work is based on data from eROSITA, the soft X-ray instrument aboard SRG, a joint Russian-German science mission supported by the Russian Space Agency (Roskosmos), in the interests of the Russian Academy of Sciences represented by its Space Research Institute (IKI), and the Deutsches Zentrum für Luft- und Raumfahrt (DLR). The SRG spacecraft was built by Lavochkin Association (NPOL) and its subcontractors, and is operated by NPOL with support from the Max Planck Institute for Extraterrestrial Physics (MPE).

The development and construction of the eROSITA X-ray instrument was led by MPE, with contributions from the Dr. Karl Remeis Observatory Bamberg \& ECAP (FAU Erlangen-Nuernberg), the University of Hamburg Observatory, the Leibniz Institute for Astrophysics Potsdam (AIP), and the Institute for Astronomy and Astrophysics of the University of Tübingen, with the support of DLR and the Max Planck Society. The Argelander Institute for Astronomy of the University of Bonn and the Ludwig Maximilians Universität Munich also participated in the science preparation for eROSITA. 

The eROSITA data shown here were processed using the eSASS/NRTA software system developed by the German eROSITA consortium.

Based on observations made with ESO Telescopes at the La Silla Paranal Observatory under programme IDs 177.A-3016, 177.A-3017, 177.A-3018 and 179.A-2004, and on data products produced by the KiDS consortium. The KiDS production team acknowledges support from: Deutsche Forschungsgemeinschaft, ERC, NOVA and NWO-M grants; Target; the University of Padova, and the University Federico II (Naples).

\bibliographystyle{aa}
\bibliography{biblio}

\appendix

\section{Optical Cluster Lensing}
\label{appendix:gglensing}
\begin{figure}
    \centering
    \includegraphics[width=\columnwidth]{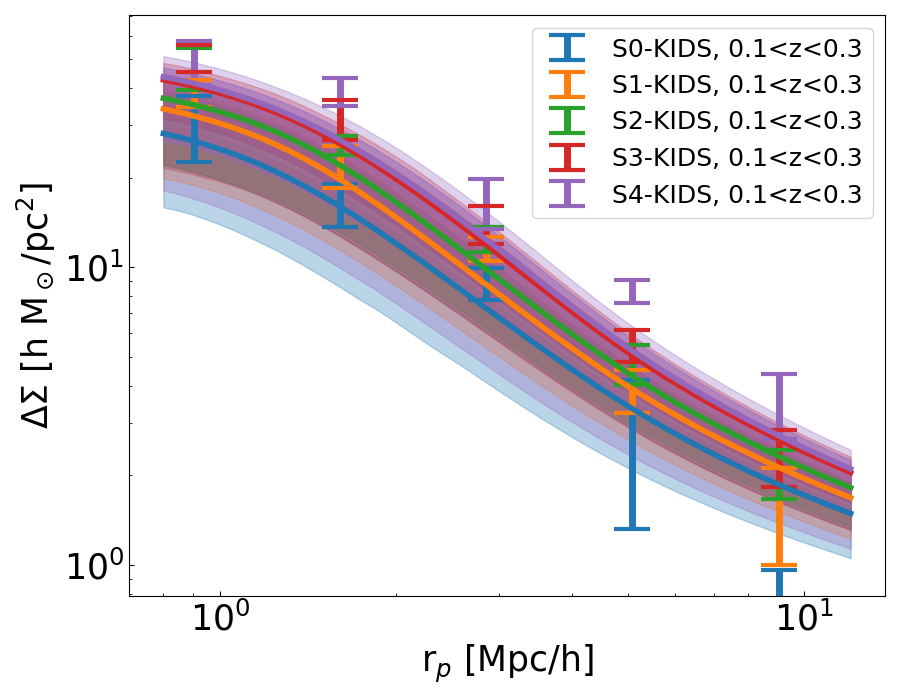}
    \caption{Excess surface density as a function of radial separation measured around eRASS1 clusters with KiDS data. The prediction from our best-fit HOD model is shown by the solid lines. The shaded areas denote the 1-$\sigma$ and 2-$\sigma$ confidence in the prediction. Each colour denotes one of the volume-limited samples (see Table \ref{tab:z_sample}).}
    \label{fig:deltasigma}
\end{figure}

As a consistency check, we test our HOD model against a measurement of the excess surface density from optical cluster lensing. The latter is a cross-correlation between clusters acting as foreground lenses at redshift $z_l$ and the shape of background galaxies acting as sources at redshift $z_s$ \citep[][]{Bartelmann2001PhR...340..291B}. The principle is the same as galaxy-galaxy lensing \citep[][]{Mandelbaum2005MNRAS_gglensing}, but in our case, the lens is a cluster and not a galaxy. We use the galaxy shapes from the Kilo Degree Survey (KiDS) DR4 \citep[][]{Kuijken2019A&AKiDS_dr4}. 
The cross-correlation with lensing encodes the difference between the mean density within a given aperture and the density at the aperture location. This quantity is named excess surface density $\Delta \Sigma$ (or ESD). It is related to the galaxy-matter correlation function by Eq. \ref{eq:dsigma}:
\begin{align}
    \Sigma(R) &= \overline{\rho_m} \int \xi_{gm}(\sqrt{R^2+\pi^2})d\pi, \nonumber \\
    \Delta \Sigma &= \overline{\Sigma(<R)} - \Sigma(R) = \Sigma_{\rm crit}\gamma(R),
    \label{eq:dsigma}
\end{align}
where $\Sigma_{\rm crit}=\dfrac{c^2}{4\pi G} \dfrac{D_{\rm A}(z_s)}{D_{\rm A}(z_l)D_{\rm A}(z_s,z_l)}\dfrac{1}{(1+z_l)^2}$ is the critical surface density, which depends on the angular angular diameter distances between the lens and the source planes, and $\gamma(R)$ is the cluster shear profile. 
We measure $\Delta \Sigma$ with the publicly available \texttt{dsigma} software\footnote{\url{https://dsigma.readthedocs.io}} \citep[][]{Lange2022_dsigma}. The excess surface density is estimated according to:
\begin{equation}
    \Delta \Sigma = \frac{\sum_{ls} w_{ls} \Sigma_{\rm crit}(z_l,z_s) e_t}{\sum_{ls} w_{ls}},
    \label{eq:dsigma_measure}
\end{equation}
where $e_t$ is the source tangential ellipticity, $w_s$ is a source weight that minimizes shape noise, and $w_{ls}=\dfrac{w_s}{\Sigma^2_{\rm crit}(z_l,z_s)}$.\\
We focus on the redshift range 0.1<z<0.3. We require the redshift of background sources to be $z_s$ > $ z_l$ + 0.1. The model is evaluated at the average redshift of each cluster sample acting as a lens. The result is shown in Fig. \ref{fig:deltasigma}. It shows the measurement of KiDS data with error bars. The prediction by our HOD model is shown by the solid lines, with 1-$\sigma$ and 2-$\sigma$ uncertainty denoted by the shaded areas. Overall, we find a good agreement between the data and the prediction. We conclude that our HOD model is consistent with optical cluster lensing measurements. In future studies, accounting for mis-centering, cluster member contamination, and photo-z uncertainties, in ESD models for clusters will allow us to use this quantity to improve the HOD constraining power for clusters \citep[see][for AGN]{Comparat2023A&Aagnclustering}.

\end{document}